\documentclass[sigconf]{acmart}

\usepackage{wrapfig,stfloats}
\usepackage{subcaption}
\usepackage[utf8]{inputenc} %
\usepackage[T1]{fontenc}    %
\usepackage{hyperref}       %
\usepackage{url}            %
\usepackage{booktabs}       %
\usepackage{amsfonts}       %
\usepackage{nicefrac}       %
\usepackage{microtype}      %
\usepackage{xcolor}         %
\usepackage{graphicx}
\usepackage{xspace}
\usepackage{multirow}
\usepackage{tikz}
\usepackage{comment}
\usepackage{enumitem}
\newcommand*\circled[1]{\tikz[baseline=(char.base)]{\node[shape=circle,fill=white,draw,text=black,inner sep=1pt] (char) {\textbf{#1}};}}

\newcommand{\zl}[1]{{#1}}
\newcommand{\sys}{\texttt{LlamaTouch}\xspace}
\newcommand{\agentenv}{\texttt{AgentEnv}\xspace}

\usepackage{pifont}%
\newcommand{\xmark}{\textcolor{red}{\ding{55}}\xspace}
\newcommand{\cmark}{\textcolor{green}{\ding{51}}\xspace}

\copyrightyear{2024}
\acmYear{2024}
\setcopyright{acmlicensed}\acmConference[UIST '24]{The 37th Annual ACM Symposium on User Interface Software and Technology}{October 13--16, 2024}{Pittsburgh, PA, USA}
\acmBooktitle{The 37th Annual ACM Symposium on User Interface Software and Technology (UIST '24), October 13--16, 2024, Pittsburgh, PA, USA}
\acmDOI{10.1145/3654777.3676382}
\acmISBN{979-8-4007-0628-8/24/10}

\begin{document}

\title[\sys: A Faithful and Scalable Testbed for Mobile UI Task Automation]{\sys: A Faithful and Scalable Testbed for\\Mobile UI Task Automation}

\author{Li Zhang}
\affiliation{
\institution{State Key Laboratory of Networking and Switching Technology,\\Beijing University of Posts and Telecommunications
}
\country{}
}
\author{Shihe Wang}
\affiliation{
\institution{State Key Laboratory of Networking and Switching Technology,\\Beijing University of Posts and Telecommunications
}
\country{}
}
\author{Xianqing Jia}
\affiliation{
\institution{State Key Laboratory of Networking and Switching Technology,\\Beijing University of Posts and Telecommunications
}
\country{}
}
\author{Zhihan Zheng}
\affiliation{
\institution{State Key Laboratory of Networking and Switching Technology,\\Beijing University of Posts and Telecommunications
}
\country{}
}
\author{Yunhe Yan}
\affiliation{
\institution{State Key Laboratory of Networking and Switching Technology,\\Beijing University of Posts and Telecommunications
}
\country{}
}
\author{Longxi Gao}
\affiliation{
\institution{State Key Laboratory of Networking and Switching Technology,\\Beijing University of Posts and Telecommunications
}
\country{}
}

\author{Yuanchun Li}
\affiliation{
\institution{Institute for AI Industry Research (AIR), Tsinghua University}
\country{}
}

\author{Mengwei Xu}
\affiliation{
\institution{State Key Laboratory of Networking and Switching Technology,\\Beijing University of Posts and Telecommunications
}
\country{}
}

\renewcommand{\shortauthors}{Zhang et al.}

\begin{abstract}
The emergent large language/multimodal models facilitate the evolution of mobile agents, especially in mobile UI task automation.
However, existing evaluation approaches, which rely on human validation or established datasets to compare agent-predicted actions with predefined action sequences, are unscalable and unfaithful.
To overcome these limitations, this paper presents \sys, a testbed for on-device mobile UI task execution and faithful, scalable task evaluation.
By observing that the task execution process only transfers UI states, \sys employs a novel evaluation approach that only assesses whether an agent traverses all manually annotated, essential application/system states.
\sys comprises three key techniques:
(1) \textit{On-device task execution} that enables mobile agents to interact with realistic mobile environments for task execution.
(2) \textit{Fine-grained UI component annotation} that merges pixel-level screenshots and textual screen hierarchies to explicitly identify and precisely annotate essential UI components with a rich set of designed annotation primitives.
(3) \textit{A multi-level application state matching algorithm} that utilizes exact and fuzzy matching to accurately detect critical information in each screen, even with unpredictable UI layout/content dynamics.
\sys currently incorporates four mobile agents and 496 tasks, encompassing both tasks in the widely-used datasets and our self-constructed ones to cover more diverse mobile applications.
Evaluation results demonstrate \sys's high faithfulness of evaluation in real-world mobile environments and its better scalability than human validation.
\sys also enables easy task annotation and integration of new mobile agents.
\zl{
Code and dataset are publicly available at \url{https://github.com/LlamaTouch/LlamaTouch}.
}
\end{abstract}



\maketitle
\section{Introduction}

Mobile intelligent agents empower users to interact with their smartphones using natural languages, alleviating them from tedious and cumbersome smartphone operations.
These agents are particularly beneficial for individuals with visual or hand impairments, or in situations where using a screen is not practical (e.g., driving).
Notable mobile agents, such as Apple Siri~\cite{siri} and Google Assistant~\cite{google-assistant}, have become indispensable services on smartphones.
The recent advent of large language models (LLMs) and multimodal LLMs has facilitated researchers in building more powerful mobile agents~\cite{li2024personal,wen2023empowering,wang2023enabling,lee2023explore}.
The key capability of these agents is to comprehend user instructions in natural language and execute corresponding actions on the mobile interface, as called \textit{mobile UI task automation}, e.g., ``forward the last email from Bob to Alice''.

Despite claims of powerful task automation capabilities achieved by recent LLM-powered mobile agents, their evaluation methods are somewhat flawed.
Unlike traditional machine learning models evaluated on well-established static datasets, mobile agents need to interact with the dynamic and indeterministic states of a smartphone (e.g., network connectivity, dynamic content) as inputs.
\zl{
Additionally, mobile devices use touch and gesture-based interactions (e.g., swipe, pinch), leading to diverse and ambiguous inputs.
This variability complicates static action-based matching algorithms.
}
Therefore, simply evaluating mobile agents using deterministic smartphone states from datasets cannot uncover their true capabilities~\cite{rawles2023android,li2020mapping,sun2022meta}.

In general, there are two methods to evaluate mobile UI automation tasks, but neither achieves both high faithfulness and scalability.
(1) The most intuitive approach is to request humans to verify the completion of tasks.
However, human evaluation is difficult to reproduce~\cite{chiang2023can}, and the requisite human effort increases with the number of agents, tasks, and evaluation platforms.
(2) The most popular approach used in most prior work~\cite{rawles2023android,yan2023gpt,sun2022meta,wen2023empowering,zhan2023you,hong2023cogagent} is \textit{exact action match on established datasets}, akin to traditional machine learning evaluations.
The key idea is to ask annotators to generate a correct sequence of actions that succeed on the task as \textit{data labels}, and then compare agent-generated actions to these labels.
Although this approach allows for some error tolerance, e.g., the variations in click positions on the screen~\cite{rawles2023android}, it cannot cover all possible and ``infinite'' paths to complete a UI automation task.
Consequently, it leads to a significantly higher false negative rate.
For example, for the task \textit{``Reserve a rental car in Los Angeles from June 1st-7th, with a budget of up to \$60 per day on Expedia''}, the sequence of three filtering actions can be interchanged.
Only taking one of these execution paths as the reference may incorrectly verify a task that is essentially completed.
Moreover, LLM-powered agents are known to be able to self-correct their wrong actions~\cite{pan2023autotask}, which is critical to enhancing UI task automation capabilities, yet is impossible to evaluate in a static dataset.
These limitations are further demonstrated in $\S$\ref{sec:bkgnd-benchmarks}.

This paper presents \sys, the first testbed for evaluating mobile agents in real-world mobile environments without compromising faithfulness and scalability.
The key idea of \sys is to check the task execution trace against a few ``essential states'' identified by the annotators, rather than matching them against predefined action sequences in static traces.
For instance, the essential states for the task \textit{``open app Microsoft Excel (install if not already installed), go to login, ...''} should include
(1) the application ``Microsoft Excel'' is opened, and (2) the application is on the login page.
Other operations, like app installation, are considered non-essential and should be ignored.
During task execution, \sys enables mobile agents to retrieve only task descriptions from static datasets, while device states are directly acquired from realistic mobile devices.
Actions produced by mobile agents are directly operated on those devices, and all UI interaction data are recorded as task execution traces.
In the evaluation phase, \sys compares task execution traces with annotated essential states to determine whether a task has been completed.

To ensure faithful and scalable evaluation, \sys integrates two effective methods.
(1) \sys adopts a fine-grained labeling mechanism for essential state annotation at both the screen level and single UI component level.
It combines pixel-level screenshots and textual screen hierarchies to explicitly highlight important UI components.
With a rich set of annotation primitives provided by \sys, it reduces human effort to heuristically identify and annotate the attributes of essential states for evaluation, e.g., the text inside a textbox should be exactly matched. 
These annotated UI states are subsequently used for faithful evaluation.
(2) During evaluation, \sys employs a multi-level state matching algorithm that combines fuzzy and exact matches on diverse annotated UI states.
It uses (i) approximate screen matching, which enables \sys to adapt to dynamic mobile environments and varying screen contents, and (ii) mixed UI state matching, which detects and matches critical on-screen information.

\textbf{Dataset and testbed.}
We present a large-scale dataset with pre-annotated essential states for evaluating mobile UI automation tasks in real-world mobile environments.
This dataset includes 496 distinct tasks encompassing a wide array of popular Android applications.
We complement this dataset with an easy-to-use testbed that enables mobile agents to interact seamlessly with realistic Android environments.
This testbed provides a collection of concise, widely used APIs, ensuring compatibility with most mobile agents.
Mobile agents can be easily integrated into \sys and use our dataset to test their capabilities in mobile UI task automation in real-world scenarios.

\textbf{Evaluation.}
We implemented \sys by utilizing Google Android emulator~\cite{android-emulator} and one Google Pixel 5 smartphone as realistic Android environments.
Currently, \sys has four built-in agents, including AutoUI~\cite{zhan2023you}, AppAgent~\cite{yang2023appagent}, AutoDroid~\cite{wen2023empowering}, and CoCo-Agent~\cite{ma2024comprehensive}, along with 496 diverse tasks.
With human validation results as the ground truth for task completion, \sys achieves nearly 80\% evaluation accuracy in detecting completed tasks in real-world environments, while prior action-based evaluation methods fail to do so.
We also reveal the limitations of current mobile agents in handling tasks practically in real-world environments.

\textbf{Contributions} are summarized as follows.
\begin{itemize}
	\item We observed the weakness of high false negative rates in evaluating mobile UI task automation agents using static datasets.
    To address this, we proposed an evaluation design that only compares essential states rather than concrete action sequences.
    \item We devised a method for annotating essential states using a variety of annotation primitives.
    This approach combines visually intuitive screenshots with semantically precise view hierarchies to enable fine-grained and accurate UI component localization and annotation.
	\item We designed a novel task evaluation approach that employs both exact and fuzzy matching at various UI state levels.
	It enables faithful evaluation of mobile agents and adapts well to dynamic execution environments.
	\item We proposed \sys, the first testbed to faithfully and salably evaluate mobile UI task automation agents in real-world mobile environments.
    It comprises 496 tasks with human-annotated essential states.
    Four agents integrated in \sys demonstrate its faithfulness and scalability in UI automation task evaluation.
\end{itemize}

\begin{table}[t]
    \caption{The comparison between mobile agent benchmarks.
    \textmd{
    \sys is the first testbed designed for mobile agents driven by essential state matching.
    \sys also supports fine-grained UI-guided essential state annotation with a rich set of primitives covering a wide array of matching implementations.}
    }
    \renewcommand{\arraystretch}{1.1}
    \scalebox{0.67}{
    \begin{tabular}{cccccc}
    \hline
    \textbf{Benchmark} & \textbf{Platform} & \textbf{\begin{tabular}[c]{@{}c@{}}Real-world\\ Tasks\end{tabular}} & \textbf{\begin{tabular}[c]{@{}c@{}}Real-env\\ Task Exec\end{tabular}} & \textbf{\begin{tabular}[c]{@{}c@{}}Fine-grained\\UI Annotation\end{tabular}} & \textbf{\begin{tabular}[c]{@{}c@{}}Essential\\State Match\end{tabular}} \\ \hline
    Rico~\cite{deka2017rico} & \multirow{8}{*}{Mobile} & \cmark & \xmark & \xmark & \xmark \\
    PixelHelp~\cite{li2020mapping} &  & \cmark & \xmark & \xmark & \xmark \\
    AndroidEnv~\cite{toyama2021androidenv} &  & \xmark & \cmark & \xmark & \xmark \\
    META-GUI~\cite{sun2022meta} &  & \cmark & \xmark & \xmark & \xmark\\
    MoTIF~\cite{burns2022dataset} &  & \cmark & \xmark & \xmark & \xmark\\
    AITW~\cite{rawles2023android} &  & \cmark & \xmark & \xmark & \xmark\\
    Mobile-Env~\cite{zhang2023mobile} &  & \cmark & \cmark & \xmark & \xmark \\
    AndroidArena~\cite{xing2024understanding} &  & \cmark & \cmark & \xmark & \xmark\\ \hline
    WebArena~\cite{zhou2023webarena} & Web & \cmark & \cmark & \xmark & \cmark\\ \hline
    \textbf{\sys} & \textbf{Mobile} & \cmark & \cmark & \cmark & \cmark\\ \hline
    \end{tabular}
    }
    \footnotesize
    \label{tab:benchmark-comp}
    \vspace{-15pt}
    \end{table}

\section{Background and Motivation}

\subsection{Agents for Mobile UI Task Automation}

Mobile agents have simplified the cumbersome and dull operations on smartphones for users.
The progression of mobile agents for mobile UI task automation can be categorized into three phases.
(1)
API-based agents like Google Assistant~\cite{google-assistant} and Apple Siri~\cite{siri} interact with applications through predefined application programming interfaces.
This approach is reliable while limited in structured and predictable tasks.
(2)
Learning-based agents~\cite{rawles2023android,sun2022meta,li2020mapping,zhan2023you} utilize deep learning techniques to learn from previous mobile interaction traces, but their capabilities are still confined by their training data.
(3)
Recently, LLMs and multi-modality LLMs have revolutionized the capabilities of mobile agents~\cite{gpt4v,hong2023cogagent,wen2023empowering,wang2023enabling}.
These models, owing to their vast knowledge base, can understand complex, real-world mobile screens.
Mobile agents powered by these models can accurately interpret natural language instructions and translate them into actionable tasks on smartphone screens.
This evolution marks a significant leap in the flexibility and adaptability of mobile agents.

Mobile UI task automation agents typically operate with the following components.

\noindent \textbf{Controller} is the brain of mobile agents.
It interprets task instructions and UI contexts, and then generates actions to be executed on the current UI context.
Widely-used controllers include deep learning models tailored for specific applications~\cite{rawles2023android,sun2022meta,li2020mapping,zhan2023you}, LLMs (e.g., GPT-4, Llama)~\cite{wen2023empowering,wang2023enabling,lee2023explore}, and multi-modality LLMs (e.g., GPT-4V)~\cite{yan2023gpt,hong2023cogagent,yang2023appagent}.

\noindent \textbf{Input: UI Representation.}
Existing mobile agents take a task description and UI representations as the input of their controller.
There are two basic types of UI representations: screenshot and view hierarchy (VH).
A screenshot is a visual capture of the current screen.
A VH provides a textual tree-like structure of the UI elements present on a screen, including their properties such as type, position, and text contents.
On top of screenshots and VHs, some controllers further extract UI semantics to enhance UI understanding.
For example, Yan et al.~\cite{yan2023gpt} overlay numeric tags on top of each text and icon detected by OCR tools;
AXNav~\cite{taeb2023axnav} converts screenshots to bounding boxes and labels, making them comprehensible to LLMs.
Further processing based on VH, such as converting it to simple HTML representations, is also widely utilized~\cite{wang2023enabling,wen2023empowering}.

\noindent \textbf{Output: Action.}
The output of controllers consists of actions to be executed on the current screen, such as click, swipe, and input text.
Action parameters can be abstracted at different levels depending on the agent's design and input format.
(1) Concrete coordinates on the screen~\cite{rawles2023android,sun2022meta,hong2023cogagent,zhan2023you}: This operates as a direct interaction with the screen, similar to human operations.
(2) Icon marker~\cite{yan2023gpt,yang2023appagent}: The output target will specify a specific icon or graphical element within the UI representation.
(3) HTML index~\cite{wang2023enabling,wen2023empowering,lee2023explore}: By ingesting HTML representations, controllers will give a concrete HTML index as the action target, which matches specific elements (icons or text) on the screen.

\begin{figure}[b]
	\centering
	\begin{minipage}[t]{0.48\textwidth}
		\centering
		\includegraphics[width=1\textwidth]{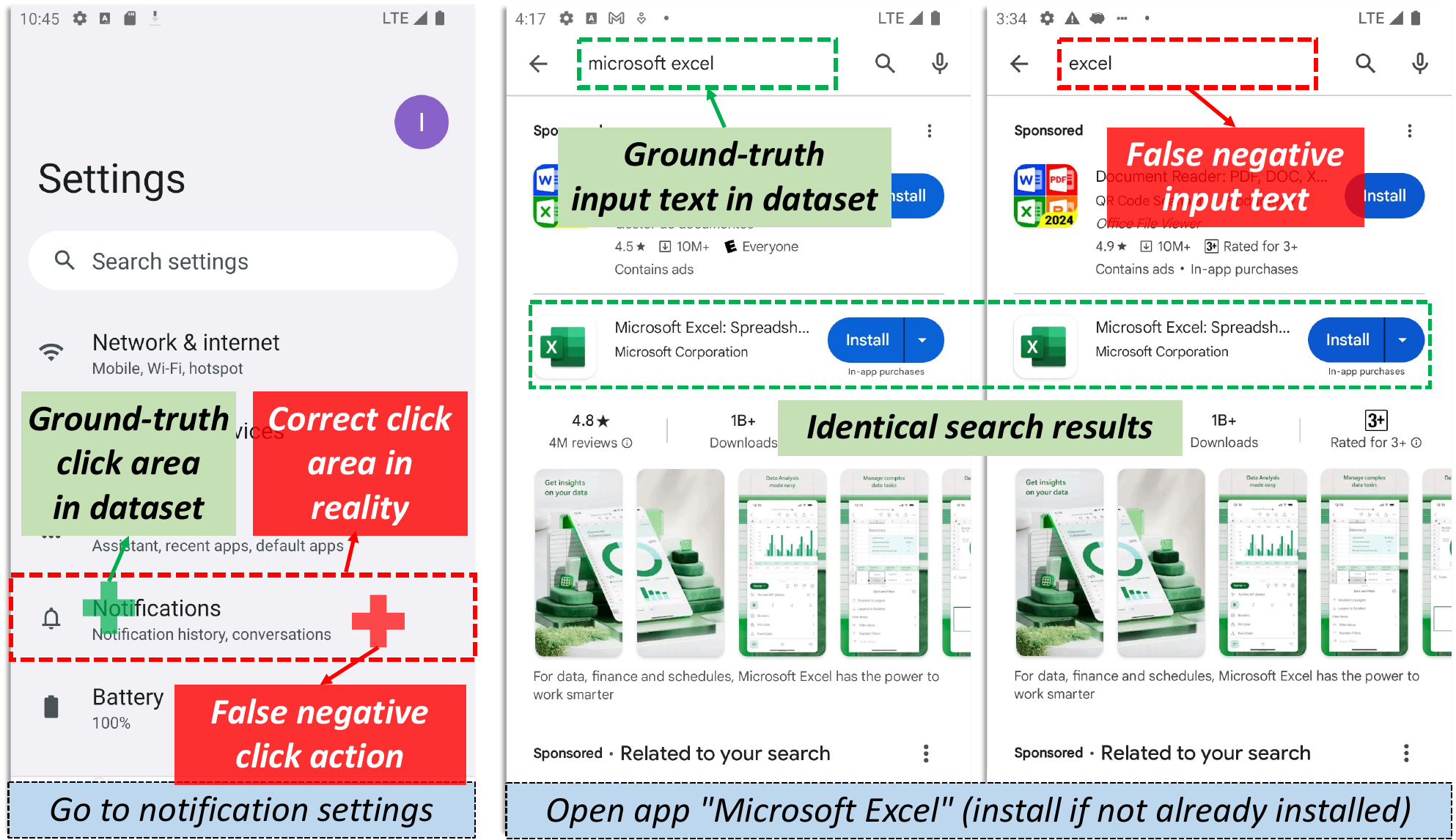}
		\subcaption{Inaccurate action match in two tasks.
		\textmd{Failed reasons: ``Left'': wrong click parameter; ``Right'': wrong input text.}}
		\label{fig:incorrect-action-match}
	\end{minipage}
	
	\hspace{10pt}
	\begin{minipage}[t]{0.48\textwidth}
		\centering
		\includegraphics[width=1\textwidth]{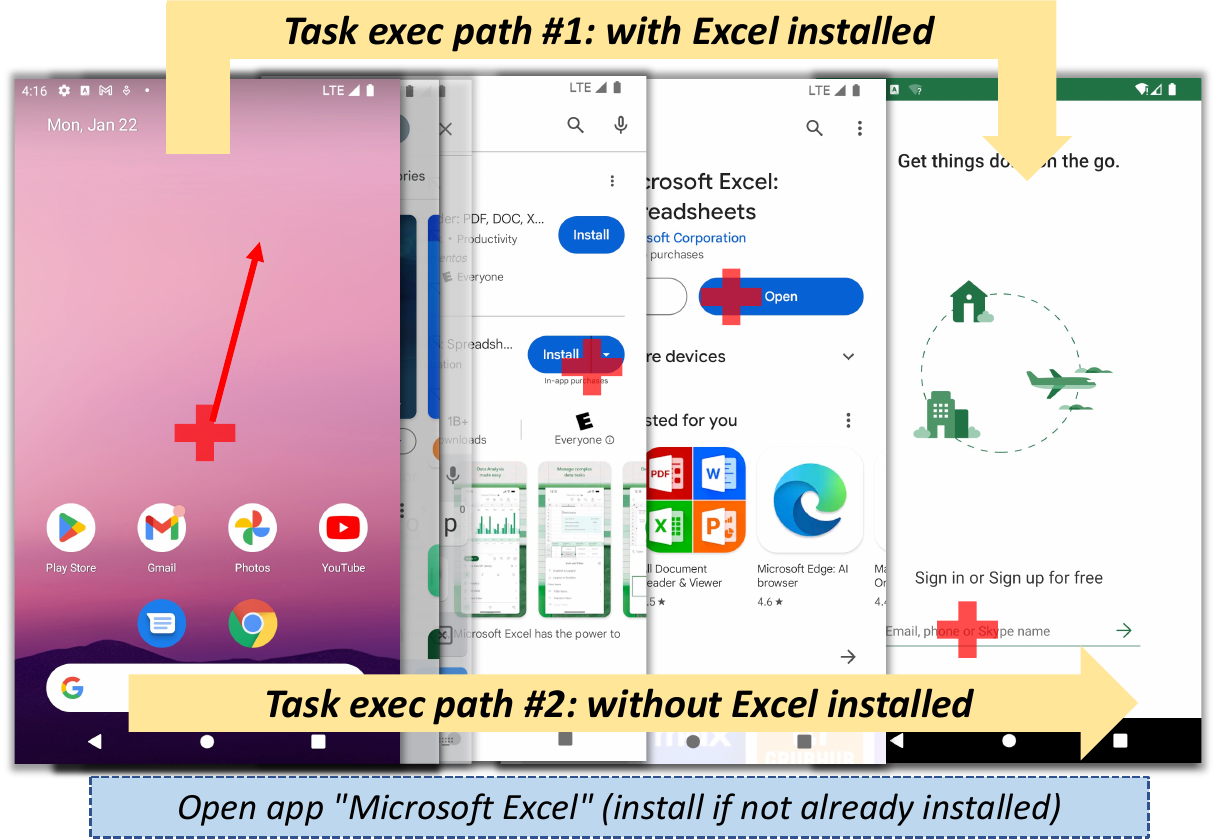}
		\subcaption{Different task execution paths lead to the same screen.}
  \vspace{-5pt}
		\label{fig:diff-task-exec-path}
	\end{minipage}
	\caption{\zl{Two major limitations of evaluating mobile UI task automation on static datasets.}}
 \vspace{-10pt}
\end{figure}

\subsection{Mobile UI Task Automation Benchmarks}
\label{sec:bkgnd-benchmarks}

\begin{figure*}[t]
	\centering
	\includegraphics[width=0.99\textwidth]{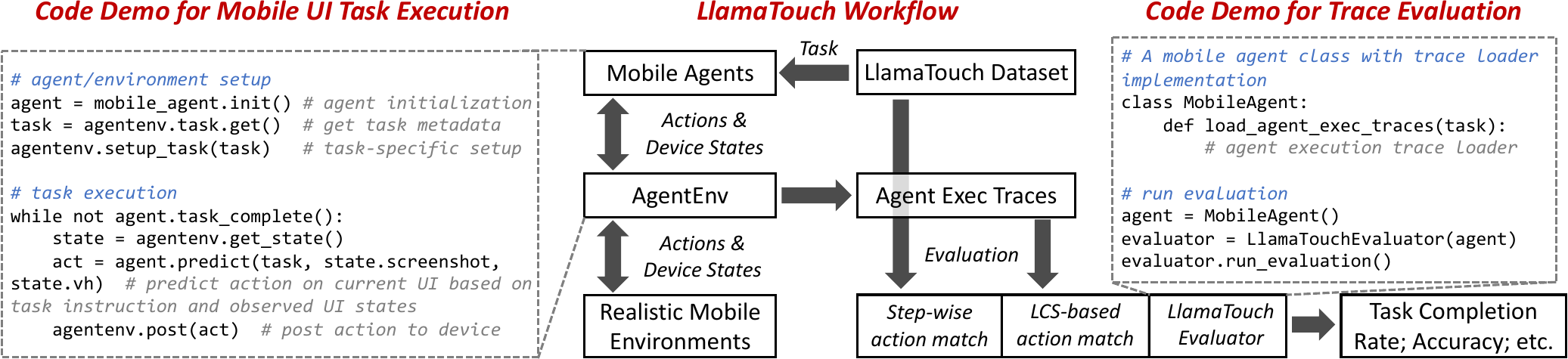}
	\caption{
		\zl{\sys workflow and code demonstrations for mobile UI task execution and trace evaluation.
	\textmd{\sys enables mobile UI task automation agents to integrate easily with \agentenv for on-device task execution with minimal programming effort.
	Agent execution traces recorded by \agentenv are used in conjunction with the \sys dataset in a separate evaluation process.}
	}}
	\label{fig:llamatouch-workflow}
\end{figure*}

As shown in Table~\ref{tab:benchmark-comp}, a variety of datasets and environments are proposed to evaluate mobile agents in UI task automation, but none of them achieve both faithfulness and scalability.

Some work such as Rico~\cite{deka2017rico}, PixelHelp~\cite{li2020mapping}, and AITW~\cite{rawles2023android}, provides static datasets with task descriptions, UI representations, and actions sequences.
Mobile agents predict concrete actions on static UI representations, which are compared with the ground-truth actions in the datasets.
While this approach is straightforward, it is insufficient to reveal the performance of mobile agents for two reasons.
\textit{(1) Inaccurate exact action match.}
Functionally correct actions may be deemed incorrect due to different action parameters.
Figure~\ref{fig:incorrect-action-match} illustrates two cases.
First, for click actions, the clickable area defined in the dataset may be narrow, whereas in reality, the area might encompass the entire UI component (marked with the red bounding box).
Second, non-identical text inputs can lead to the same correct search result (marked with green bounding boxes) in most search tasks.
This issue has been observed in previous literature but remains unsolved~\cite{yan2023gpt,rawles2023android}.
\textit{(2) Lack of tolerance for different execution paths.}
In real-world environments, a task can usually be completed in various paths based on different device/application states, as shown in Figure~\ref{fig:diff-task-exec-path}.
However, predefined datasets might only provide one deterministic path for reference, leading to inaccurate evaluation.

There is also other work that enables agent execution in real-world environments, such as AndroidEnv~\cite{toyama2021androidenv} and Mobile-Env~\cite{zhang2023mobile}.
However, they do not inherently support essential state match during end-to-end task execution, therefore compromising evaluation accuracy.
AndroidArena~\cite{xing2024understanding} observed the weakness of step-wise action match on static datasets: 
it does not fully tolerate redundant actions in task execution paths.
They proposed a subsequence-based action match, where a task is treated as completed if it contains the ground-truth action sequence as its subsequence.
We take AndroidArena as a baseline in $\S$\ref{sec:eval-e2e} to compare its evaluation accuracy with \sys.
WebArena~\cite{zhou2023webarena} provides a realistic playground for web agents.
It uses essential states to evaluate task completion (e.g., the final result should be or should include some key information).
\sys differs from WebArena on the mobile platform in both essential state annotation and evaluation process:
(i) \sys combines visual screenshots with textual VHs of the same screens for fine-grained and precise UI component identification.
(ii) \sys uses a richer set of primitives to comprehensively annotate essential UI states and faithfully evaluate them even with high screen content dynamics.

Human validation is usually used to validate whether a UI automation task is completed~\cite{rawles2023android,taeb2023axnav}.
However, the cost of human validation is too high, scaling poorly to multiple tasks, agents, and mobile devices.
\sys ensures high scalability as with evaluating on static datasets while preserving faithfulness similar to human validation.

\section{\sys Design}

\zl{
Figure~\ref{fig:llamatouch-workflow} shows the workflow of using \sys for on-device task execution and trace evaluation, using the well-constructed \sys dataset.
}
Compared to previous evaluation approaches, \sys exhibits the following benefits.

$\bullet$ Practical on-device mobile UI task execution ($\S$\ref{sec:design-on-device-exec}).
Most previous evaluation methods are simulated on mobile UI interaction datasets.
\sys, on the other hand, enables mobile agents to operate in realistic mobile environments for UI task automation, revealing their true capabilities in real-world scenarios.

$\bullet$ Fine-grained essential application state annotation ($\S$\ref{sec:design-UI-match}). 
By observing UI automation task execution transfers states of essential UI elements within an application, \sys enables annotators to explicitly annotate essential application states that should be detected and matched for task completion.
This approach avoids the previous static evaluation methods' focus on the determinism of traces, reducing the probability of false negatives in evaluation.

$\bullet$ Faithful and scalable task evaluation ($\S$\ref{sec:design-scalability}).
\sys evaluates the performance of mobile agents by comparing their task execution traces captured in real-world mobile environments with annotated essential states.
By combining exact and fuzzy matching algorithms on different application states, \sys achieves faithful task evaluation without losing scalability.

\subsection{On-device Mobile UI Task Execution}\label{sec:design-on-device-exec}
\begin{table*}[t]
\caption{A set of primitives used in essential state annotation and trace evaluation.}
\vspace{-10pt}
\scalebox{0.85}{
    \begin{tabular}{c|c|c|c|l}
    \hline
    \textbf{Match Type}          & \textbf{State Type}       & \textbf{Primitive} & \textbf{Keyword}                                                                                                                   & \multicolumn{1}{c}{\textbf{Use Case}}                                                                                                                                        \\ \hline
    \multirow{2}{*}{Fuzzy match} & \multirow{5}{*}{UI state} & Screen info        & \texttt{fuzzy\textless{}-1\textgreater{}}                                                                                                   & Verify if the contents on two screens are approximately identical.                                                                                                                                                      \\ \cline{3-5} 
                                 &                           & Textbox            & \texttt{fuzzy\textless{}n\textgreater{}}                                                                                                    & \begin{tabular}[c]{@{}l@{}}Verify if the content of the target textbox is semantically similar\\ to the content of the original textbox\textless{}n\textgreater{} in the ground-truth UI.\end{tabular}                                        \\ \cline{1-1} \cline{3-5} 
    \multirow{4}{*}{Exact match} &                           & Activity           & \texttt{activity}                                                                                                                           & \begin{tabular}[c]{@{}l@{}}A coarse-grained approach to determine whether two UIs represent\\ the same functional screen in an application.\end{tabular}                                                                           \\ \cline{3-5} 
                                 &                           & UI component       & \begin{tabular}[c]{@{}c@{}}\texttt{exact\textless{}n\textgreater{},}\\\texttt{exclude\textless{}n\textgreater{}}\end{tabular}                                                                                                    & \begin{tabular}[c]{@{}l@{}}Verify if the UI component is exactly identical to the UI component\textless{}n\textgreater{},\\or if it does not occur, in the ground-truth UI.\end{tabular}                                           \\ \cline{2-5} 
                                 & System state              & (Un)installation   & \begin{tabular}[c]{@{}c@{}}\texttt{installed\textless{}app\textgreater{},}\\\texttt{uninstalled\textless{}app\textgreater{}}\end{tabular} & \begin{tabular}[c]{@{}l@{}}Verify if the target application named "app" has been successfully\\ installed or uninstalled.\end{tabular}                                                                                   \\ \cline{2-5} 
                                 & Action                    & Action             & \begin{tabular}[c]{@{}c@{}}\texttt{click\textless{}n\textgreater{},}\\\texttt{type\textless{}input\_text\textgreater{}}\end{tabular}                & Verify if two actions and their parameters are identical.                                                                                                                                                            \\ \hline
    \end{tabular}
}
\label{tab:primitives}
\end{table*}

\begin{figure}[t]
	\centering					
	\includegraphics[width=0.44\textwidth]{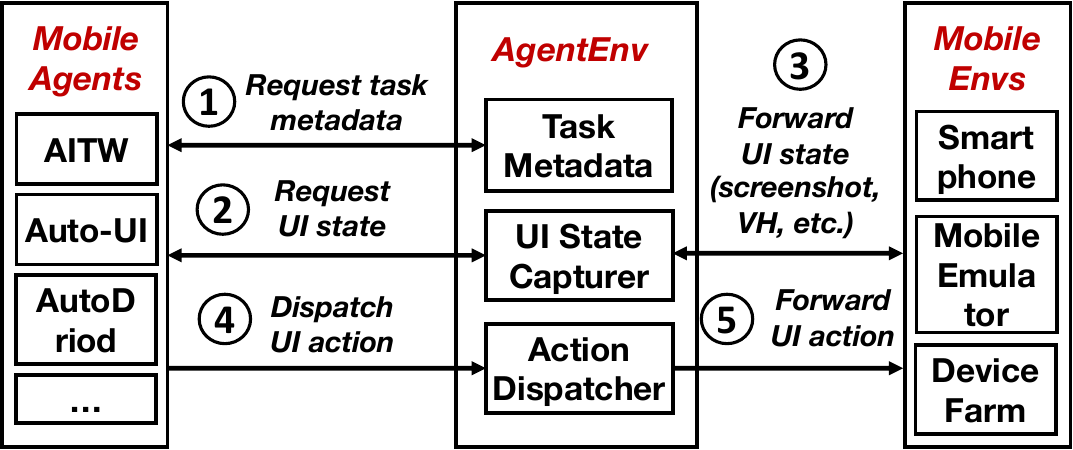}
	\caption{Interaction between mobile agents, \agentenv, and real-world mobile environment in the on-device mobile UI task execution process.}
	\label{fig:device-exec}
 \vspace{-15pt}
\end{figure}

Evaluating mobile agents on predefined, deterministic traces, as discussed in Section~\ref{sec:bkgnd-benchmarks}, depicts significant inaccuracy.
\sys empowers on-device mobile UI task execution to reveal the real capabilities of mobile agents.
To simplify this process, we propose \agentenv, a bridge between existing mobile agents and realistic mobile environments (e.g., smartphones, Android emulators, and cloud device farms).
With \agentenv, mobile agents execute tasks on real-world mobile environments by following the processes shown in Figure~\ref{fig:device-exec}.
\textbf{\circled{1}} Mobile agents request a task instruction from \agentenv.
The instruction comes from the \sys dataset.
\textbf{\circled{2}} With the task instruction, mobile agents request UI representations (e.g., screenshot, VH) from \agentenv, which are then \textbf{\circled{3}} forwarded to mobile devices.
\textbf{\circled{4}} Taking the task instruction and UI representations as inputs, mobile agents predict an action to be performed on the current UI and dispatch the predicted action to \agentenv.
\textbf{\circled{5}} \agentenv forwards and executes the agent-predicted action to mobile environments.
The processes from \textbf{\circled{2}} to \textbf{\circled{5}} are repeated until mobile agents consider the task completed.
During task execution, all UI representations and corresponding actions are captured as task execution traces for further evaluation in $\S$\ref{sec:design-scalability}.
These UI representations include (1) pixel-level screenshots; (2) textual screen VHs; (3) activity names of the application in the foreground of each screen; and (4) actions performed on each screen.
Essential system states, such as the list of installed applications, will also be recorded for faithful mobile agent evaluation.
\zl{
The left-hand side of Figure~\ref{fig:llamatouch-workflow} demonstrates the code implementation of the interaction between mobile agents and \agentenv.
Appendix~\ref{appendix:apis} presents details of the APIs provided by \sys for mobile agent integration.
}

\subsection{Essential Application State Annotation}\label{sec:design-UI-match}

Two major cases demonstrated in $\S$\ref{sec:bkgnd-benchmarks} highlight why exact action match on predefined action sequences is unfaithful.
First, it demands that the agent-generated actions and their parameters match exactly with those in the dataset.
Second, it takes a fixed UI interaction sequence provided by the dataset as the reference, making it unable to evaluate alternative task execution paths.

\textbf{Insight: The task execution process transfers identifiable application states.}
During the execution of a UI automation task, the application states change, and some of these states can be explicitly represented by UI components.
Even if the task execution paths differ, there are overlaps in the essential application states.
We can utilize these overlapping application states to determine whether a task achieves some milestones or is completed.
As the example shown in Figure~\ref{fig:diff-task-exec-path}, by identifying the intent of the task ``open app Microsoft Excel (install if not already installed), go to login, ...'', it contains two potential essential states for evaluation:
(1) the application ``Microsoft Excel'' is opened; 
and (2) the application is located at the login page.
Other actions, like detecting whether the application is installed, can be omitted during evaluation.

\begin{figure*}[t]
	\centering
	\includegraphics[width=0.9\textwidth]{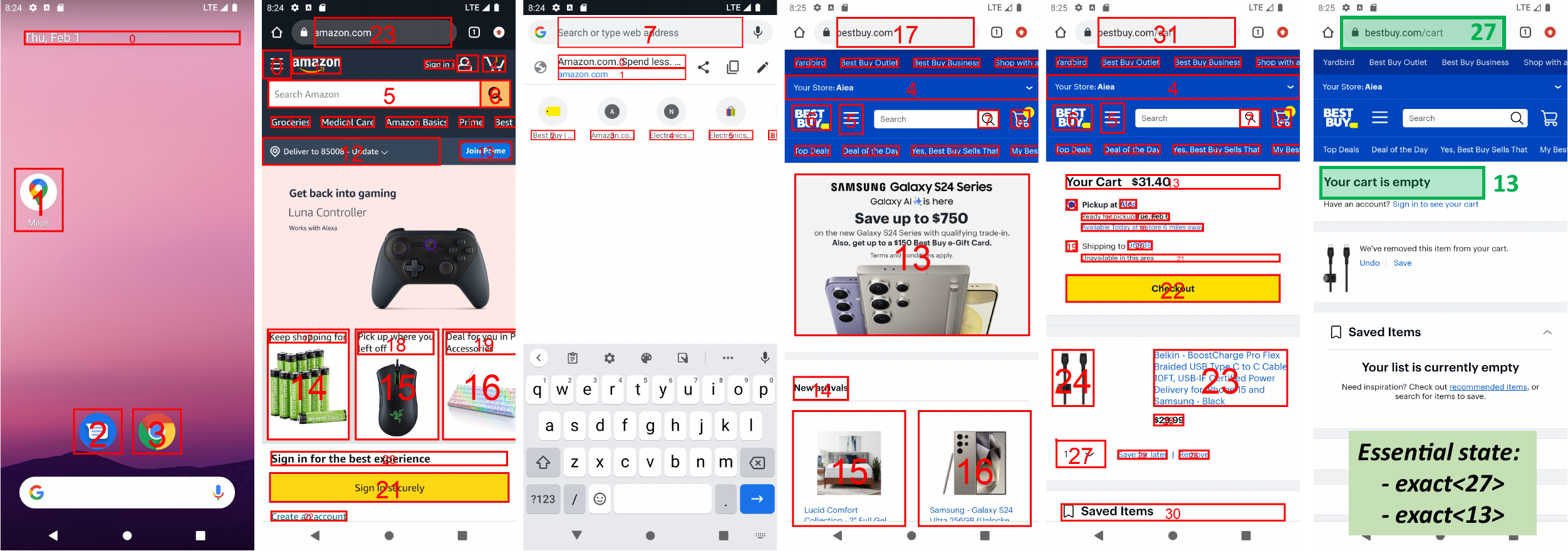}
	\caption{Annotated essential states for the task ``empty the shopping cart on bestbuy'' in the last UI representation: two textboxes with the \texttt{exact} keyword.
	The essential states represent the application state after task execution: ``the shopping cart on bestbuy is empty''.
	}
	\label{fig:state-annotation}
 \vspace{-10pt}
\end{figure*}

To achieve this, essential states should be accurately identified and annotated.
However, simply annotating application states at the whole UI representation level (e.g., an entire screenshot) and comparing screen-level similarity~\cite{feiz2022understanding} is too coarse-grained and may lead to inaccurate matching.
For example, the screen contents of a web-shopping application could be subtly different due to nondeterministic swiping gestures or dynamically loaded contents across different executions.
For accurate and efficient essential state annotation, \sys breaks the whole pixel-level UI representation into separate UI components.
This is achieved by simplifying the textual VH of each screen, which precisely expresses the attributes of every UI element, to extract important, visible UI components.
These extracted UI components are combined with visually intuitive screenshots to provide precise overlayed bounding boxes and unique identifiers to annotators.
Figure~\ref{fig:state-annotation} shows an example of VH-enhanced screenshots in a task provided to annotators.

\textbf{Annotation primitives.}
\sys incorporates a list of primitives for essential state annotation.
These primitives represent essential information on the screen and indicate how this information should be matched.
Annotators are responsible for clearly identifying and using primitives to represent application states that are informative and deterministic for validating task execution results.
Table~\ref{tab:primitives} comprehensively shows these primitives and their use cases.
Currently, \sys incorporates six types of primitives and nine keywords for annotation.
These primitives can be divided into three types according to the application state type they represent.

$\bullet$ \textit{UI state.}
The purpose of UI state annotation is to extract and compare whether two screens contain identical or similar information.
As a screen may contain different types of components, \sys uses four major primitives to cover the comparison logic of all these UI components.
(1) \textit{Screen info} and (2) \textit{textbox} are used to compare whether the whole screen or a dedicated textbox has similar contents.
\textit{Screen info} is primarily used for checking whether two UI representations are within the same screen (i.e., the same page of an application).
\textit{Textbox} primitive is used to annotate textboxes that may contain dynamic contents, such as a search box in a web-shopping application or the URL of a website.
(3) The functionality of \textit{activity} is like \textit{screen info}: it can be used to approximately detect whether two UI representations are on the same screen of an application.
(4) \textit{UI component} is used to deterministically annotate the state of a UI component, including textbox, button, image, etc.
For example, if the content of a textbox should be matched, this textbox should be annotated with the \texttt{exact} keyword.
States that require certain UI components not to occur on the screen can be annotated with the \texttt{exclude} keyword.

$\bullet$ \textit{System state.}
System state annotation in \sys is primarily inspired by tasks from the previously built dataset: AITW~\cite{rawles2023android}.
For example, a task like ``install app YouTube Kids'' can be detected directly using shell commands to access system states (e.g., \texttt{pm list packages}), without involving complex UI states.
Currently, \sys supports programmatically checking application installation status.
\zl{
The keywords can be easily extended and customized to detect other system or application states.
}

$\bullet$ \textit{Action.}
Although the principle of annotation is to detect application state transfer during task execution.
Sometimes, action on a specific UI representation is necessary to validate agent behavior, especially when there are not enough state identifiers shown in the UI.
\sys provides \texttt{click} and \texttt{type} keywords that cover the most common actions.

\textbf{Case study.}
Essential state annotation can be divided into the following processes.
First, for each UI representation, \sys overlays all functional UI components with numeric markers on the screenshot as shown in Figure~\ref{fig:state-annotation}.
This is done by extracting the precise metadata of each UI component from textual VH.
Annotators will then identify the essential states that should be checked during evaluation to ensure the task is completed.
\sys simplifies this process by only requiring annotators to explicitly identify what UI element should be matched and what annotation primitive should be used.
For example, after emptying the shopping cart on bestbuy, the screen will display a textbox with the content ``Your cart is empty''.
This textbox should be treated as an essential state, which is highlighted in a bounding box with numeric ID 13 in the last screen of Figure~\ref{fig:state-annotation}, as it represents the state after task completion.
As we anticipate the content of the textbox should be exactly matched, the annotated keyword is \textit{exact<13>}.
The keyword \textit{exact<27>} is used to validate whether two screens are both in the shopping cart of bestbuy.
All annotations, along with the UI representations and task descriptions, construct the essential state-powered dataset.
These essential states are used by \sys for faithful task evaluation.

\subsection{Faithful and Scalable Task Evaluation}\label{sec:design-scalability}

Annotated essential states ($\S$\ref{sec:design-UI-match}) and captured task execution traces ($\S$\ref{sec:design-on-device-exec}) are jointly used for evaluation in \sys.
\sys iterates through essential states to determine whether a task execution trace sequentially matches all annotated states.
If so, the task is deemed completed.
To achieve faithful evaluation, the most significant challenge lies in how to ensure the task execution trace matches the essential states.
Task execution traces captured from real-world mobile environments may contain dynamic screen content;
it is vital to adapt the annotated essential states from a static dataset to varying screens captured from the real world, while achieving precise matches on only critical information.
To address this problem, \sys employs a multi-level state matching algorithm, which combines fuzzy match and exact match on both the entire screen and separated UI components to ensure faithful evaluation.
The algorithm first approximately matches two screens according to their UI representations and activities.
Then, within two matched screens, it compares each annotated essential state with iterated UI components in the target screen based on their annotated primitives.
A task is considered completed only when all annotated essential states have matching counterparts.
The matching philosophy of all annotation primitives is as follows.

\textbf{Approximate screen match} is used to ensure two screens are on the same functionality page of an application, even under high screen content dynamics.
This is the preliminary process for in-screen exact/fuzzy content match.
To achieve this, \sys utilizes two annotation primitives proposed in $\S$\ref{sec:design-UI-match}: \textit{activity} and \textit{screen info}.

\textit{1. Application activity match.}
Activity represents an entry point for users to interact with the application~\cite{android-activity}.
This is a clear identifier to indicate whether two screens are in the same application.
Screens with distinct functionalities typically have unique activity names, even if they are within the same application.
For example, the activity of the main settings page is ``com.android.settings.Settings'', while the Wi-Fi settings page has the activity ``com.wifiadmin.settings. WifiSettingsActivity''.
Typically, exact application activity match acts as a foundational filtering process to identify whether two screens are on the same page. 
However, some specific application design philosophies may cause different functional pages of an application to contain the same activity name \cite{single-activity}.
Under these circumstances, \sys involves \textit{screen info} for accurate screen match.

\textit{2. Fuzzy screen info match} is utilized when the activity name cannot differentiate between functional pages of an application.
To better compare two screens, it is crucial to extract only the critical information from the screen.
In summary, \sys simplifies the textual screen VH to a simple HTML representation, as in prior work~\cite{wen2023empowering,wang2023enabling}, while preserving the types of every UI component.
\sys compares two simplified HTML representations of the screens using the cosine similarity of their sentence embeddings.
Two screens are deemed similar when their cosine similarity exceeds a predefined threshold, e.g., 0.85 in our experiments.
The fuzzy screen info match design helps \sys maintain faithfulness during evaluation when dealing with dynamic screen contents.

\begin{figure*}[th]
	\centering
	\includegraphics[width=0.98\textwidth]{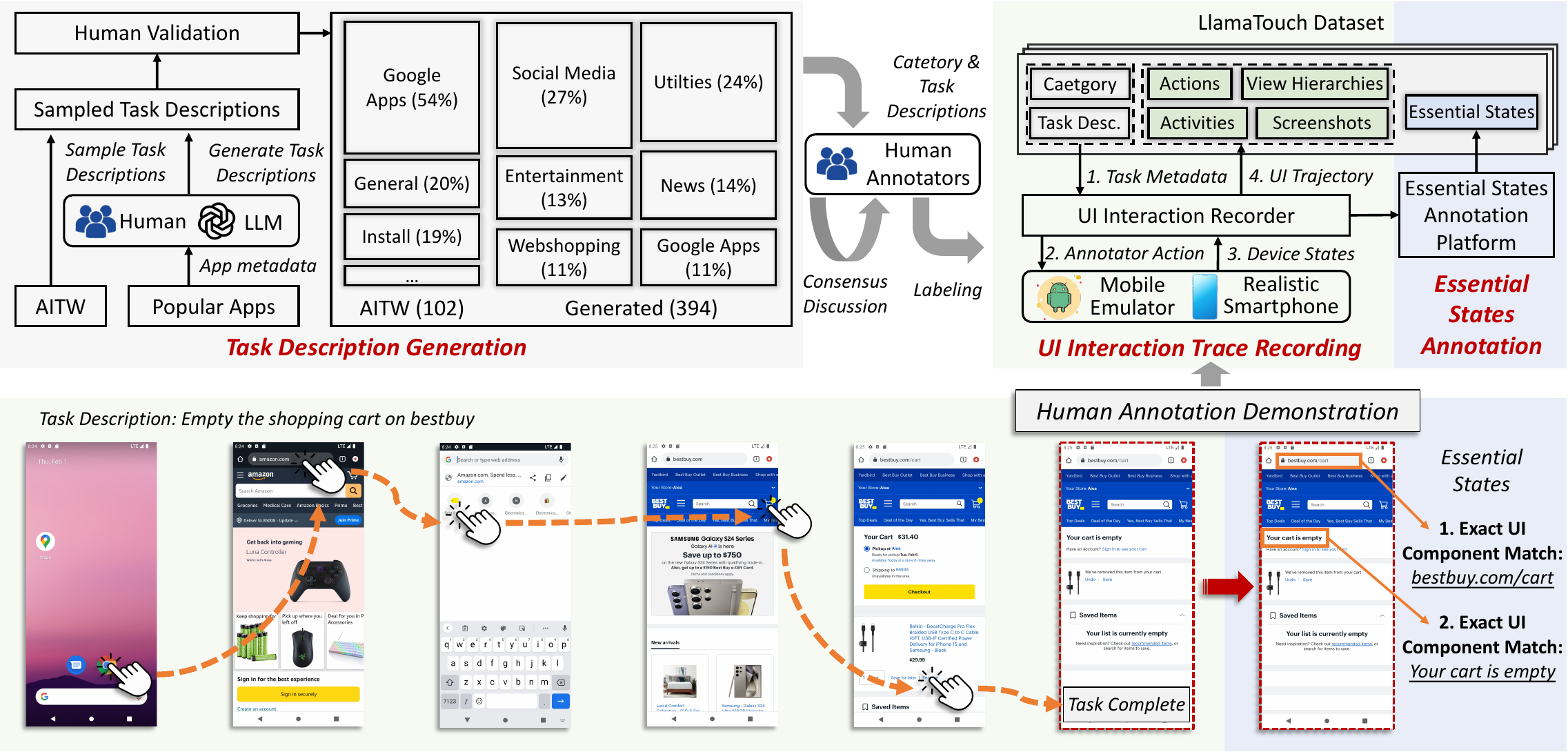}
	\caption{\zl{Workflow for constructing the \sys dataset.
	\textmd{
	Generally, the workflow has three independent processes.
	(1) Generate task descriptions by sampling from previous datasets or constructing new ones through humans or LLMs.
	(2) Record UI interaction traces according to task descriptions.
	(3) Annotate essential states atop UI interaction traces.}}}
	\label{fig:dataset-construction}
\end{figure*}

\textbf{Mixed UI state match} will be applied to matched screens after the approximate screen match process.
A mix of annotated UI states will be checked in this phase, including both fuzzy match and exact match on UI components, actions, and system states.

\textit{1. Exact UI component match} 
requires that an annotated UI component be identical in two screens being compared, including all their attributes such as \textit{class}, \textit{text}, and \textit{selected} in VH.
This is especially useful for evaluating the content of a textbox, the status of a button (e.g., checked or not), or a selectable icon on the screen.
Given a target UI component to be exactly matched, \sys will iterate through nodes in the VH of the matched screen until a matching UI component is found.
Exact UI component match fails if no matching node is found on available screens.
In our dataset, exact UI component match accounts for 51\% (698 out of 1,379) of annotated essential states.

\textit{2. Fuzzy textbox match} is crucial for comparing the content inside a textbox on the screen, especially when the content may be slightly different.
Semantically similar search keywords with the same intent that comply with a specific task description should be matched, as they will lead to the same results.
For example, Figure~\ref{fig:diff-task-exec-path} shows searching for ``Microsoft Excel'' and ``Excel'' in the Google Play Store both display the target application.
\sys extracts the content of the annotated textbox and then compares the text with nodes in the matched screen using the same approach as in \textit{fuzzy screen info match}.

\textit{3. Action match.}
Although the initiative of \sys is to detect state transfer during task execution,
there are still cases for evaluating concrete actions performed on the screen, such as clicking a specific UI component or typing the correct captcha.
\sys directly compares the actions and their parameters of the annotated actions with their counterparts on the matched screens.
\zl{
Different from prior studies~\cite{rawles2023android}, \sys uses XPath~\cite{xpath} of target UI components extracted from screen VHs as parameters for click actions.
This approach provides more tolerance in the action space compared to using precise coordinates.
}

\textit{4. System state match} is usually more efficient and accurate than merely comparing UI states for specific tasks that involve deterministic system states, e.g., installed applications.
\sys currently supports checking whether an application is installed or not.
Such system states are recorded during task execution on real-world mobile devices ($\S$\ref{sec:design-on-device-exec}).
\sys will check whether the annotated system state is identical to that of the last screen in the task execution trace, which is also recorded by \agentenv.

Through the multi-level state matching algorithm, \sys achieves high evaluation accuracy on real-world task execution traces, while preserving the scalability of evaluating on static datasets.
The above evaluation logic is well encapsulated into the \sys evaluator.
To use it for evaluation, an agent only needs to define how to load the task execution traces for each task, as shown in the right side of Figure~\ref{fig:llamatouch-workflow}, 
The evaluator will automatically conduct the evaluation and report metrics such as task completion rate.
Experiments in $\S$\ref{sec:eval} show the faithful evaluation of \sys.

\zl{\section{\sys Dataset}}
\label{sec:dataset}

\zl{\subsection{Dataset Construction}}

\zl{
The \sys dataset consists of a combination of tasks from the previous AITW dataset~\cite{rawles2023android} and self-generated tasks that involve diverse categories and popular applications.
The inclusion of new tasks in currently popular applications, which have not been covered in previous literature, aims to properly assess the generality and real capabilities of mobile agents.
Figure~\ref{fig:dataset-construction} demonstrates the workflow for constructing the dataset, where each data sample undergoes the following processes.
}

\zl{
\noindent \textbf{Generate task descriptions.}
Task descriptions are generated by both humans and LLMs using app metadata (e.g., app names, categories, descriptions) from popular apps in the Google Play Store.
For tasks in the AITW dataset, we sample parts of them after deduplicating similar task descriptions.
The sampled task descriptions are then validated by humans to avoid duplication, infeasibility, and high complexities that may far exceed the capabilities of human and mobile agents.
After this process, we sampled 102 tasks from the AITW dataset among 26 unique apps and 394 newly generated tasks among 46 unique apps.
The generated descriptions cover a variety of application categories such as utilities (e.g., Zoom, Expedia), social media (e.g., Discord, Instagram), and web shopping (e.g., Walmart, Amazon).
The left-hand side of Figure~\ref{fig:dataset-construction} shows the proportion of these categories.
}

\zl{
We then employ six human annotators, all of whom are authors of this study and experts in smartphone usage, to generate data samples in \sys through the following two independent annotation stages.
The image on the lower side of Figure~\ref{fig:dataset-construction} illustrates the outputs of the human annotation process.
}

\zl{
\noindent \textbf{Record UI interaction traces.}
The validated task descriptions are used as guidance for recording UI interaction traces.
Tasks sampled from the AITW dataset are also required to go through this process as they lack view hierarchies.
Given a task description, human annotators interact with mobile apps through our developed \textit{UI Interaction Recorder}.
This tool is built on top of mobile emulators or realistic smartphones and displays the graphical user interface to users, allowing them to operate it like normal smartphones.
Specifically, human annotators are asked to complete a task in the simplest manner and to avoid redundant operations, as in~\cite{rawles2023android}.
As shown in Figure~\ref{fig:dataset-construction}, the task ``Empty the shopping cart on bestbuy'' requires five continuous click actions to complete.
The recorder captures actions, VHs, activities, and screenshots, which collectively form the UI trajectory.
}

\zl{
\noindent \textbf{Annotate essential states.}
The recorded UI interaction traces, along with task descriptions, are then processed by human annotators to identify essential states.
To help annotators better understand the application state transformation and simplify the annotation process, we developed an essential state annotation system.
This system displays the entire UI interaction trace, with potentially significant UI components shown with numeric indices in each screenshot.
Annotators are asked to identify the most significant, identifiable states that represent key milestones during task completion.
They annotate these states (their numeric indices) with the proper essential state primitives we proposed in $\S$\ref{sec:design-UI-match}.
These essential states, together with recorded UI interaction traces, form the final \sys dataset.
}

\zl{
During the annotation process, to ensure data reliability, anything that a single human annotator cannot decide on is annotated based on a consensus reached by three or more annotators.
Overall, this dataset includes 496 tasks, covering 57 unique Android applications with diverse task complexities.
}

\zl{
\subsection{Dataset Statistics}
}

\zl{
In this section, we present statistics of the \sys dataset.
Table~\ref{tab:dataset-stats} quantifies task complexities by showing the average steps (actions) required to complete a task.
Tasks from AITW~\cite{rawles2023android} are slightly more complex than those generated by \sys, with average steps of 7.35 versus 5.67, respectively.
Overall, the average number of steps to complete a task is nearly 7, with task complexities ranging from 2 to 42 steps.
}

\begin{table}[h]
    \centering
    \caption{\sys dataset statistics and task complexities measured by the average steps (actions) to complete a task.}
    \scalebox{0.9}{
    \renewcommand{\arraystretch}{1.1}
    \begin{tabular}{c|c|c|c}
    \hline
    \textbf{Category}       & \textbf{\# Task} & \textbf{\# Apps} & \textbf{Avg. Steps} \\ \hline
    \textbf{AITW~\cite{rawles2023android}}     & 102               & 26               & 7.35 (2-19)              \\
    \textbf{Generated} & 394               & 46               & 5.67 (3-42)              \\ \hline
    \textbf{Total}               & 496               & 57               & 7.01 (2-42)              \\ \hline
    \end{tabular}
    \footnotesize
    }
    \label{tab:dataset-stats}
\end{table}

\zl{
We further analyze action types in the ground-truth dataset.
Table~\ref{tab:action-stats} shows the statistics of actions contributing to the dataset.
In summary, all 496 tasks involve \textit{click} actions.
Out of these, 292 tasks involve \textit{swipe} actions, and 147 tasks involve \textit{input} actions.
A few tasks also use press \textit{home} and \textit{back} for navigation.
By dividing the total number of actions by the number of tasks involving these actions, we observe an average of more than 4 \textit{click} actions per data sample.
The average number of all other actions is less than 2.
}

\begin{table}[]
	\caption{Action statistics in \sys dataset.}
	\scalebox{0.9}{
	\begin{tabular}{c|c|c|c}
	\hline
	\textbf{Action}     & \textbf{\# Action} & \textbf{\# Tasks W/ Action} & \textbf{Mean/Stddev} \\ \hline
	\textbf{Click}      & 2,192              & 496                         & 4.42/2.51            \\
	\textbf{Swipe}      & 376                & 292                         & 1.29/0.9             \\
	\textbf{Input Text} & 173                & 147                         & 1.18/0.55            \\
	\textbf{Press Home} & 44                 & 43                          & 1.02/0.15            \\
	\textbf{Press Back} & 6                  & 5                           & 1.2/0.45             \\ \hline
	\end{tabular}
	}
	\footnotesize
	\label{tab:action-stats}
	\end{table}

\begin{table}[t]
    \centering
    \caption{\zl{Statistics of annotated essential states.}}
    \scalebox{0.85}{
    \begin{tabular}{c|c|ccc}
    \hline
    \textbf{Type}                                                          & \textbf{Essential State (ES)} & \textbf{\# ES} & \textbf{\# Tasks W/ ES} & \textbf{Mean/Stddev} \\ \hline
    \multirow{6}{*}{\begin{tabular}[c]{@{}c@{}}Exact\\ Match\end{tabular}} & UI Component                  & 698            & 418                     & 1.67/0.92            \\
                                                                           & Activity                      & 490            & 442                     & 1.11/0.32            \\
                                                                           & Action: Click                 & 98             & 93                      & 1.05/0.23            \\
                                                                           & Action: Type                  & 1              & 1                       & 1/0                  \\
                                                                           & System: Install               & 7              & 7                       & 1/0                  \\
                                                                           & System: Uninstall             & 3              & 3                       & 1/0                  \\ \hline
    \multirow{2}{*}{\begin{tabular}[c]{@{}c@{}}Fuzzy\\ Match\end{tabular}} & UI Component                  & 51             & 44                      & 1.16/0.43            \\
                                                                           & Screen Info                   & 31             & 31                      & 1/0                  \\ \hline
    \end{tabular}
    \footnotesize
    }
    \label{tab:primitive-stats}
\end{table}

\zl{
Statistics in Table~\ref{tab:primitive-stats} detail the annotated essential states.
We find that the essential states of 441 tasks can be presented on only one UI representation, indicating that most tasks check only the final states after task completion.
There are 53 tasks and 2 tasks with essential states presented on two and three UI representations, respectively.
Among the annotated essential states, \textit{exact UI component match} accounts for 51\% of all essential states, followed by \textit{exact activity match} at 36\%.
More than 84\% of tasks (418 out of 496) have at least one of these two types of essential states.
Fuzzy match is also important for the evaluation process: 44 tasks have 51 \textit{fuzzy UI component match} in total;
\textit{fuzzy screen info match} occur in 31 tasks.
In summary, all types of essential states construct the dataset, which significantly contributes to the faithful evaluation of mobile UI task automation, as we will show in the following section.
}

\section{Evaluation}
\label{sec:eval}

\subsection{Experiment Setup}

\noindent \textbf{Mobile environments.}
\sys primarily utilizes an x86-64 Android emulator~\cite{android-emulator} as the mobile environment for task execution.
The Android emulator is configured with Android 12 (API level 31) with Google Play Services deploying on a Linux server running Ubuntu 18.04 OS.
Tasks involving applications that are not compatible with the Android emulator (e.g., Snapchat) are tested on a real Google Pixel 5 device with Android 14 OS.

\zl{
\noindent \textbf{Task setup.}
In this study, tasks are set up in two stages.
First, we construct an Android system image for the Android emulator, which provides a concrete environment where partial task states are prepared.
This ensures a controlled and replicable starting point for the tasks.
Second, we develop setup scripts using the Android UIAutomator2 library~\cite{uiautomator2} to manage complex setup processes with specific constraints (e.g., require manual login or run on real-world smartphones).
These scripts automate the initialization of the environment, ensuring tasks are ready for execution with minimal manual intervention.
}

\noindent \textbf{Mobile agents.}
We selected four mobile agents that cover diverse types of brains, including supervised learning models, LLM, and large multi-modality models.
\begin{itemize}
    \item Auto-UI~\cite{zhan2023you} employs a multimodal transformer model based on BLIP-2~\cite{li2023blip} and T5~\cite{raffel2020exploring} variants as the brain for decision-making.
    It takes pixel-level screenshots and task descriptions as input.
    \item AutoDroid~\cite{wen2023empowering} uses LLMs for device control.
    It first takes textual screen VH and simplifies it to an explicit, readable HTML representation.
    The simplified HTML representation is then processed by LLMs to generate the corresponding action.
    We selected GPT-4 (gpt-4-0125-preview)~\cite{achiam2023gpt} as the backend of AutoDroid.
    \item AppAgent~\cite{yang2023appagent} utilizes GPT-4V~\cite{gpt4v} to comprehend the screenshots of mobile devices and then dispatch controlling commands to \sys.
    In our evaluation, we used the AppAgent version without document pre-exploration for simplicity.
    \item CoCo-Agent~\cite{ma2024comprehensive} uses LLaVA~\cite{liu2023visual} as the brain.
    It is trained on a subset of the original AITW dataset.
\end{itemize}

\noindent \textbf{Evaluation methodologies.}
We used the following methodologies for evaluating whether an agent completes a task.

\begin{itemize}
    \item \textit{Step-wise action match} is widely-utilized in evaluating mobile agents on well-established datasets~\cite{rawles2023android,sun2022meta,wen2023empowering,zhan2023you}.
    It compares the agent-generated action sequences in real-world environments with ground-truth action sequences in the dataset.
    When two action sequences are identical, the task is treated as completed.
    We utilize the action match algorithm in AITW~\cite{rawles2023android}: two actions are matched only if they have identical action types and parameters.
    \item \textit{Longest common subsequence (LCS)-based action match} is proposed in AndroidArena~\cite{xing2024understanding}.
    It is built upon step-wise action match by tolerating redundant actions between ground-truth ones.
    Two action sequences are matched when the execution action sequence contains the ground-truth sequence as the subsequence.
    \item \textit{\sys} that compares agent execution traces with our annotated essential states to check whether a task is completed.
    \item\textit{Human} validates whether a task is completed based on the agent execution traces.
    Humans are instructed not only to focus on the action (and its parameter) on each UI representation, but also the whole application state transfer during task execution.
    The results of human validation are treated as ground truth for comparing the accuracy of the three evaluation approaches.
\end{itemize}

\zl{
\subsection{{Metrics}}
}

\zl{
We primarily compare (1) the end-to-end task completion rate (TCR) of different mobile agents and (2) the accuracy of different evaluation approaches, using human validation results as the ground truth.
}

\zl{
\noindent \textbf{End-to-end TCR.}
Executing tasks in realistic mobile environments results in two outcomes: task completion and task non-completion.
In step-wise action match, a task is considered completed only when the two action sequences are identical.
LCS-based action match assesses whether a ground-truth action sequence is a subsequence of the agent-generated action sequence; if so, the task is evaluated as completed.
In \sys, a task is considered completed when the agent execution trace passes through all essential states in the ground-truth dataset, using the proposed algorithm in $\S$\ref{sec:design-scalability}.
The end-to-end TCR of a certain evaluation method is calculated as the number of tasks evaluated completed divided by the total number of tasks in the dataset.
}

\zl{
\noindent \textbf{Evaluation accuracy.}
In this study, we use human validation results on agent execution traces as the ground truth for task completion.
Assume $N$ denotes the ground-truth dataset encompassing all tasks.
For task $n$ in $N$, we use $H_n$ to represent the human validation result based on the agent execution trace, where $H_n$ can be either ``True'' or ``False''.
When evaluating the same agent execution trace using a specific evaluation method, the outcome is represented by $E_n$, which can also be ``True'' or ``False''.
Therefore, the accuracy formula is defined as:
\begin{equation}
    \text{accuracy} = \frac{\sum_{n \in N} \left( \delta_{E_n, H_n} \right)}{|N|}
\end{equation}
The formulation indicates that a task is accurately evaluated only when the evaluation result $E_n$ matches the human validation result $H_n$.
Consequently, the accuracy of an evaluation approach is defined as the proportion of correctly evaluated tasks to the total number of tasks in the dataset $|N|$.
}

\subsection{Task Completion Rate and Accuracy}
\label{sec:eval-e2e}

Table~\ref{tab:eval-e2e-acc} presents the end-to-end TCR and accuracy of different evaluation designs.
All three approaches achieve more than 90\% accuracy on average.
However, the step-wise action match and LCS-based action match fall short in recognizing tasks correctly executed by mobile agents, resulting in nearly 0\% TCR.
The high accuracy rates of these two approaches are attributed to the large portion of incomplete tasks; only 6\% of tasks are deemed completed upon human validation.
Compared to action match on static datasets, the TCR evaluated by \sys is 8.67\%, which closely aligns with the results of human validation.

To demonstrate \sys's effectiveness in addressing the issue of false negative results pervasive in previous evaluation methods, we focused on tasks considered successfully completed by human evaluation, excluding all incomplete tasks.
Table~\ref{tab:eval-e2e-acc-success} displays the number of tasks completed by agents and the evaluation accuracy for these tasks.
Overall, agents successfully completed 30 tasks on average.
Among these tasks, both step-wise action match and LCS-based action match achieve no more than 0.1\% evaluation accuracy.
This indicates they are unable to faithfully evaluate tasks executed in real-world environments using static datasets, primarily because a single static action sequence is difficult to match with multiple possible paths to task completion in real-world environments.
Notably, \sys exhibits an average accuracy of 79\% in validating these task execution traces, significantly reducing the percentage of false negative cases observed with other evaluation methods.
The findings uncover \sys's proficiency in evaluating UI automation tasks under real-world settings.

\begin{table}[t]
    \centering
    \caption{End-to-end task completion rate (TCR \%) and accuracy (Acc. \%) of different evaluation approaches of \underline{all tasks}.}
    \scalebox{0.86}{
    \renewcommand{\arraystretch}{1.1}
    \begin{tabular}{cccccccc}
    \hline
    \multirow{2}{*}{\textbf{\begin{tabular}[c]{@{}c@{}}Mobile\\ Agent\end{tabular}}}& \multicolumn{2}{c}{\textbf{\begin{tabular}[c]{@{}c@{}}Step-wise\\ action match\end{tabular}}} & \multicolumn{2}{c}{\textbf{\begin{tabular}[c]{@{}c@{}}LCS action\\ match\end{tabular}}} & \multicolumn{2}{c}{\textbf{\sys}}                      & \textbf{Human}            \\ \cline{2-8} 
                                                                                    & \textbf{TCR}                                  & \textbf{Acc.}                                 & \textbf{TCR}                               & \textbf{Acc.}                              & \textbf{TCR}             & \textbf{Acc.}               & \textbf{TCR}              \\ \hline
    \textbf{AutoUI}                                                                 &  0.00                                         &  98.18                                        &  0.00                                      &  98.18                                     &  4.44	                 &  96.57                      &  1.82                     \\
    \textbf{AutoDroid}                                                              &  0.00                                         &  85.98                                        &  0.00                                      &  85.98                                     &  14.84	                 &  91.87                      &  14.02                    \\
    \textbf{AppAgent}                                                               &  0.00                                         &  93.33                                        &  0.61                                      &  93.13                                     &  10.91	                 &  94.95                      &  6.67                     \\
    \textbf{CoCo-Agent}                                                             &  0.00                                         &  97.97                                        &  0.00                                      &  97.97                                     &  4.47	                 &  96.34                      &  2.03                     \\ \hline
    \textbf{Average}                                                                & \multicolumn{1}{c}{0.00}                      & \multicolumn{1}{c}{93.86}                     & \multicolumn{1}{c}{0.15}                   & \multicolumn{1}{c}{93.81}                  & \multicolumn{1}{c}{8.67} & \multicolumn{1}{c}{94.93}   & \multicolumn{1}{c}{6.14}  \\ \hline
    \end{tabular}
    \footnotesize
    }
    \label{tab:eval-e2e-acc}
\end{table}

\begin{table}[t]
    \centering
    \caption{Accuracy (Acc. \%) of different evaluation approaches among \underline{all successful tasks in human validation.}}
    \scalebox{0.86}{
    \renewcommand{\arraystretch}{1.1}
    \begin{tabular}{cccccccc}
    \hline
    \multirow{2}{*}{\textbf{\begin{tabular}[c]{@{}c@{}}Mobile\\ Agent\end{tabular}}} & \multicolumn{1}{c}{\textbf{\begin{tabular}[c]{@{}c@{}}Step-wise\\ action match\end{tabular}}} & \multicolumn{1}{c}{\textbf{\begin{tabular}[c]{@{}c@{}}LCS action\\ match\end{tabular}}} & \multicolumn{1}{c}{\textbf{\sys}}      & \textbf{Human}            \\ \cline{2-5} 
                                                                                     & \textbf{Acc.}                & \textbf{Acc.}    & \textbf{Acc.}       & \textbf{\# success} \\ \hline
    \textbf{AutoUI}                                                                  &  0.00                            &  0.00                &  77.78                  &   9             \\
    \textbf{AutoDroid}                                                               &  0.00                            &  0.00                &  73.91                  &   69              \\
    \textbf{AppAgent}                                                                &  0.00                            &  3.03                &  93.94                  &   33            \\
    \textbf{CoCo-Agent}                                                              &  0.00                            &  0.00                &  70.00                  &   10          \\ \hline
    \textbf{Average}                                                                 &  0.00                            &  0.76                &  78.91                  &   30      \\ \hline
    \end{tabular}
    \footnotesize
    }
    \label{tab:eval-e2e-acc-success}
\end{table}

\subsection{Ablation Study}
\label{sec:eval-ablation}

\sys achieves high accuracy when evaluating agent execution traces in real-world environments, attributed to the annotation and implementation of various essential state primitives.
In this section, we evaluate the effectiveness of two types of match designs in \sys: fuzzy match and exact match.
We aggregate the execution traces of all mobile agents evaluated in $\S$\ref{sec:eval-e2e}.
We evaluate the system by first disabling exact match and fuzzy match separately, and then enabling different primitives one at a time.
We present the results for (1) all tasks, (2) tasks in AITW~\cite{rawles2023android}, and (3) new tasks generated in \sys.
The results for task completion rate and accuracy are shown in Table~\ref{tab:eval-ablation}.

\textbf{Exact match} ensures key information on two screens is identical.
As shown in the results, exact match significantly contributes to improving evaluation accuracy.
For example, without all exact match primitives, \sys's evaluation accuracy on all tasks drops from 95\% to 19\%.
Among all exact match primitives, \textit{activity} match greatly improves evaluation accuracy.
This is due to its simple yet efficient ability to locate functional application screens and the large proportion of annotated activity primitives, comprising 35\% of all primitives in our dataset.
\textit{UI component match} is also necessary for faithful evaluation in \sys, as it is typically with screen location primitives such as \textit{activity} to detect critical information within the matched screen.
Exact match for \textit{action} and \textit{system state} only improves accuracy slightly because only a few tasks are annotated with these primitives.

\textbf{Fuzzy match} does not show a notable improvement in evaluation accuracy, despite the presence of 30 screen-level fuzzy match primitives and 52 textbox fuzzy match primitives in the dataset.
Few tasks with these primitives can be completed by the four agents we evaluated, leading to only a slight improvement.
Although primitives for fuzzy match are not well explored, they play an important role in dealing with screen content or UI layout dynamics in the real world.
With more performant mobile agents in the future, they will be further explored and evaluated.

\begin{table}[t]
    \centering
    \caption{Ablation study on different essential state primitives.}
    \scalebox{0.8}{
    \begin{tabular}{l|cc|cc|cc}
    \hline
    \multicolumn{1}{c|}{\multirow{2}{*}{\textbf{Evaluation design}}} & \multicolumn{2}{c|}{\textbf{All tasks}}                                & \multicolumn{2}{c|}{\textbf{AITW}}                                   & \multicolumn{2}{c}{\textbf{Generated}}                            \\ \cline{2-7} 
    \multicolumn{1}{c|}{}                                            & \multicolumn{1}{c}{\textbf{TCR}} & \multicolumn{1}{c|}{\textbf{Acc.}} & \multicolumn{1}{c}{\textbf{TCR}} & \multicolumn{1}{c|}{\textbf{Acc.}} & \multicolumn{1}{c}{\textbf{TCR}} & \multicolumn{1}{c}{\textbf{Acc.}} \\ \hline
    \textbf{Complete \sys}             &       8.67            &           94.93       &         17.46       &        89.43       &            6.38     &       96.36         \\ \hline
    \textbf{\sys W/O exact match}         &       86.77           &           18.85       &         82.55       &        29.75       &            87.87    &       16.02    \\
    + activity exact match                   &       23.21           &           81.60       &         41.05       &        68.29       &            18.58    &       85.06    \\
    + action exact match               &       86.62           &           19.01       &         81.81       &        30.49       &            87.87    &       16.02            \\
    + UI component exact match             &       15.86           &           88.15       &         40.30       &        68.55       &            9.51     &       93.24  \\
    + system state exact match      &       85.41           &           20.22       &         75.91       &        36.40       &            87.87    &       16.02              \\ \hline
    \textbf{\sys W/O fuzzy match}     &       10.54           &           93.26       &         23.36       &        84.02       &            7.21     &       95.66          \\
    + screen-level fuzzy match              &       10.24           &           93.36       &         22.13       &        84.76       &            7.15     &       95.60      \\
    + textbox fuzzy match               &       8.97            &           94.83       &         18.69       &        88.69       &            6.45     &       96.43         \\ \hline
    \end{tabular}
    }
    \footnotesize
    \label{tab:eval-ablation}
\end{table}

\subsection{Absolute Capabilities of Mobile Agents}

In this section, we present the absolute capabilities of different mobile agents in mobile UI automation tasks.
First, we categorize tasks according to their sources: AITW and our self-generated dataset.
This aims to determine whether an agent that has previously learned on a well-established dataset can adapt to new applications/tasks.
Second, we categorize tasks according to their complexities, measured by the number of steps required to complete a task in the ground-truth dataset.
Tasks are further divided into three difficulty levels: easy (steps $\le$ 4), medium (4 $<$ steps $\leq$ 8), and high (step $>$ 8).

\textbf{Performance across datasets.}
Among the four evaluated agents, AutoUI and CoCo-Agent were previously trained on AITW.
AutoDroid and AppAgent directly invoke GPT-4 and GPT-4V, respectively.
We separately show the end-to-end TCR of all agents on tasks in AITW and \sys's generated tasks to demonstrate their performance generalization to new scenarios (i.e., new tasks and new applications).
Results in Table~\ref{tab:eval-abs-agents-dataset} indicate that AutoDroid and AppAgent also achieve higher TCR in AITW tasks compared to \sys's generated tasks.
This is mainly attributed to the task complexity gap between the two datasets.
AutoUI and CoCo-Agent achieve a 12.75\% TCR on tasks in AITW.
However, for generated tasks in \sys, both perform poorly, with only 2.29\% and 2.31\% TCR, respectively.
The TCR gap between datasets reveals their lack of capability to adapt to previously unseen tasks.
Considering the vast number of real-world applications, mobile agents with strong generalization capabilities to unseen scenarios are more competitive.

\begin{table}[t]
    \centering
    \caption{The comparison of task completion rate categorized by task sources (i.e., from AITW and \sys generated tasks).}
    \scalebox{0.9}{
    \begin{tabular}{c|ccc}
    \hline
    \multirow{2}{*}{\textbf{Agent}} & \multicolumn{3}{c}{\textbf{End-to-end TCR}}           \\ \cline{2-4} 
                                    & \textbf{Overall} & \textbf{AITW} & \textbf{Generated} \\ \hline
    \textbf{Auto-UI}                &       4.44	      &	        12.75	       &        	2.29	          \\ \hline
    \textbf{AutoDroid}              &       14.84	      &	        22.77	       &        	12.79	          \\ \hline
    \textbf{AppAgent}               &       10.91	      &	        21.57	       &        	8.14              \\ \hline
    \textbf{CoCo-Agent}             &       4.47	      &	        12.75	       &        	2.31              \\ \hline
    \end{tabular}
    }
    \footnotesize
    \label{tab:eval-abs-agents-dataset}
\end{table}

\textbf{Performance under different task complexity.}
We categorize tasks into three difficulty levels based on the number of steps required to complete them in the datasets.
The results are shown in Table~\ref{tab:eval-abs-agents}.
Generally, mobile agents can better complete simple tasks that require fewer steps.
There is a significant drop in TCR when task complexity increases.
AutoDroid outperforms the other three agents for tasks at all difficulty levels.
AppAgent achieves a TCR similar to AutoDroid across all tasks.
We believe that the knowledgeable GPT-4/4V can better interpret task descriptions in natural language and pixel-level or textual UI representations.
Therefore, agents based on GPT-4/4V achieve high TCR.
However, there is still considerable room for mobile agents to improve their capabilities in mobile UI task automation.

\begin{table}[t]
    \centering
    \caption{The comparison of task completion rate between mobile agents.
    \textmd{
    Tasks are categorized into different difficulty levels according to the number of steps required to finish it in the ground-truth dataset.}
    }
    \scalebox{0.9}{
    \begin{tabular}{c|cccc}
    \hline
    \multirow{2}{*}{\textbf{Agent}} & \multicolumn{4}{c}{\textbf{End-to-end TCR}}                                                                                             \\ \cline{2-5} 
                                    & \multicolumn{1}{c}{\textbf{Overall}} & \textbf{Steps$\leq$4} & \textbf{4$<$Steps$\leq$8} & \textbf{Steps$>$8} \\ \hline
    \textbf{Auto-UI}                & \multicolumn{1}{c}{4.44}             &       4.95            &         4.19              &            4.82    \\ \hline
    \textbf{AutoDroid}              & \multicolumn{1}{c}{14.84}            &       27.00           &         12.94             &            7.32    \\ \hline
    \textbf{AppAgent}               & \multicolumn{1}{c}{10.91}            &       16.83           &         10.97             &            3.61    \\ \hline
    \textbf{CoCo-Agent}             & \multicolumn{1}{c}{4.47}             &       7.92            &         3.91              &            2.41    \\ \hline
    \end{tabular}
    }
    \footnotesize
    \label{tab:eval-abs-agents}
\end{table}

\section{Limitations of \sys}

\textbf{Supporting WebView-based apps.}
One limitation of \sys is that its evaluation process requires Android's VH to empower approximate UI layout matching and accurate UI component matching ($\S$\ref{sec:design-UI-match}).
Some Android applications built with WebView~\cite{android-webview} are unable to access their VH, preventing \sys from evaluating agent performance on these applications.
However, only a minor number of applications are built with WebView, inspiring \sys's solution in utilizing VH.
When evaluating WebView-based apps are required, more advantaged techniques, such as screen similarity detection~\cite{wu2023webui,wu2023never}, OCR/model-powered screen element recognition~\cite{bunian2021vins,xie2020uied}, should be incorporated to retrofit \sys.
We will consider this in future work.

\textbf{Biases in identifying essential states.}
\sys requires human or other autonomous agents, e.g., GPT-4V, to identify and annotate essential states on predefined UI interaction traces.
This process, however, could involve biases and leads to inaccuracy due to potentially limited knowledge in application execution and unseen circumstances.
For example, given a task ``Delete YouTube in the Google Play Store'', the predefined traces might explicitly navigate to the app page of ``YouTube'' and click the ``Uninstall'' button, which might be annotated as one essential state during task execution.
However, they may ignore that this application might be not installed at all.
In this case, there will be no explicit action of ``clicking the uninstall button'' while this task is still completed.
Such biases might result in low accuracy in checking task completion rates.
However, we think this limitation could be eliminated and refined by involving expert reviewing.
Furthermore, for every single task description, involving diverse essential states on different potential task execution paths could better enhance the robustness of \sys's evaluation design.

\section{Conclusion}
In this work, we proposed \sys, the first testbed for evaluating mobile agents with both faithfulness and scalability in mobile UI task automation.
\sys enables mobile agents to be tested on realistic mobile environments.
At the evaluation stage, it matches the task execution traces with annotated essential states.
\sys tolerates different task execution paths and dynamic execution environments, significantly reducing false negative results that occurred in previous evaluation approaches.
It achieves high evaluation accuracy, which is comparable to human validation, while preserving the scalability of evaluating on static datasets.
By conducting task execution to real-world mobile devices, we also reveal the limited capabilities of current mobile agents in mobile UI automation tasks.

\begin{acks}
This work was supported by National Key R\&D Program of China (No.2021ZD0113001), NSFC (No.62102045 and No.62272261), and China Institute of IoT (Wuxi).
\end{acks}

\bibliography{agents-ref}


\begin{thebibliography}{39}


\ifx \showCODEN    \undefined \def \showCODEN     #1{\unskip}     \fi
\ifx \showDOI      \undefined \def \showDOI       #1{#1}\fi
\ifx \showISBNx    \undefined \def \showISBNx     #1{\unskip}     \fi
\ifx \showISBNxiii \undefined \def \showISBNxiii  #1{\unskip}     \fi
\ifx \showISSN     \undefined \def \showISSN      #1{\unskip}     \fi
\ifx \showLCCN     \undefined \def \showLCCN      #1{\unskip}     \fi
\ifx \shownote     \undefined \def \shownote      #1{#1}          \fi
\ifx \showarticletitle \undefined \def \showarticletitle #1{#1}   \fi
\ifx \showURL      \undefined \def \showURL       {\relax}        \fi
\providecommand\bibfield[2]{#2}
\providecommand\bibinfo[2]{#2}
\providecommand\natexlab[1]{#1}
\providecommand\showeprint[2][]{arXiv:#2}

\bibitem[xpa(2017)]%
        {xpath}
 \bibinfo{year}{2017}\natexlab{}.
\newblock \bibinfo{title}{XML Path Language (XPath) 3.1}.
\newblock \bibinfo{howpublished}{\url{https://www.w3.org/TR/xpath-31/}}.
\newblock


\bibitem[sin(2018)]%
        {single-activity}
 \bibinfo{year}{2018}\natexlab{}.
\newblock \bibinfo{title}{Single activity: Why, when, and how (Android Dev Summit '18)}.
\newblock \bibinfo{howpublished}{\url{https://www.youtube.com/watch?v=2k8x8V77CrU}}.
\newblock


\bibitem[and(2024)]%
        {android-activity}
 \bibinfo{year}{2024}\natexlab{}.
\newblock \bibinfo{title}{Activity | Android Developers}.
\newblock \bibinfo{howpublished}{\url{https://developer.android.com/reference/android/app/Activity}}.
\newblock


\bibitem[uia(2024)]%
        {uiautomator2}
 \bibinfo{year}{2024}\natexlab{}.
\newblock \bibinfo{title}{Android UIAutomator2}.
\newblock \bibinfo{howpublished}{\url{https://github.com/appium/appium-uiautomator2-driver}}.
\newblock


\bibitem[Achiam et~al\mbox{.}(2023)]%
        {achiam2023gpt}
\bibfield{author}{\bibinfo{person}{Josh Achiam}, \bibinfo{person}{Steven Adler}, \bibinfo{person}{Sandhini Agarwal}, \bibinfo{person}{Lama Ahmad}, \bibinfo{person}{Ilge Akkaya}, \bibinfo{person}{Florencia~Leoni Aleman}, \bibinfo{person}{Diogo Almeida}, \bibinfo{person}{Janko Altenschmidt}, \bibinfo{person}{Sam Altman}, \bibinfo{person}{Shyamal Anadkat}, {et~al\mbox{.}}} \bibinfo{year}{2023}\natexlab{}.
\newblock \showarticletitle{Gpt-4 technical report}.
\newblock \bibinfo{journal}{\emph{arXiv preprint arXiv:2303.08774}} (\bibinfo{year}{2023}).
\newblock


\bibitem[Apple(2024)]%
        {siri}
\bibfield{author}{\bibinfo{person}{Apple}.} \bibinfo{year}{2024}\natexlab{}.
\newblock \bibinfo{title}{Siri - Apple}.
\newblock \bibinfo{howpublished}{\url{https://www.apple.com/siri/}}.
\newblock


\bibitem[Bunian et~al\mbox{.}(2021)]%
        {bunian2021vins}
\bibfield{author}{\bibinfo{person}{Sara Bunian}, \bibinfo{person}{Kai Li}, \bibinfo{person}{Chaima Jemmali}, \bibinfo{person}{Casper Harteveld}, \bibinfo{person}{Yun Fu}, {and} \bibinfo{person}{Magy~Seif Seif El-Nasr}.} \bibinfo{year}{2021}\natexlab{}.
\newblock \showarticletitle{Vins: Visual search for mobile user interface design}. In \bibinfo{booktitle}{\emph{Proceedings of the 2021 CHI Conference on Human Factors in Computing Systems}}. \bibinfo{pages}{1--14}.
\newblock


\bibitem[Burns et~al\mbox{.}(2022)]%
        {burns2022dataset}
\bibfield{author}{\bibinfo{person}{Andrea Burns}, \bibinfo{person}{Deniz Arsan}, \bibinfo{person}{Sanjna Agrawal}, \bibinfo{person}{Ranjitha Kumar}, \bibinfo{person}{Kate Saenko}, {and} \bibinfo{person}{Bryan~A Plummer}.} \bibinfo{year}{2022}\natexlab{}.
\newblock \showarticletitle{A dataset for interactive vision-language navigation with unknown command feasibility}. In \bibinfo{booktitle}{\emph{European Conference on Computer Vision}}. Springer, \bibinfo{pages}{312--328}.
\newblock


\bibitem[Chiang and Lee(2023)]%
        {chiang2023can}
\bibfield{author}{\bibinfo{person}{Cheng-Han Chiang} {and} \bibinfo{person}{Hung-yi Lee}.} \bibinfo{year}{2023}\natexlab{}.
\newblock \showarticletitle{Can Large Language Models Be an Alternative to Human Evaluations?}
\newblock \bibinfo{journal}{\emph{arXiv preprint arXiv:2305.01937}} (\bibinfo{year}{2023}).
\newblock


\bibitem[Deka et~al\mbox{.}(2017)]%
        {deka2017rico}
\bibfield{author}{\bibinfo{person}{Biplab Deka}, \bibinfo{person}{Zifeng Huang}, \bibinfo{person}{Chad Franzen}, \bibinfo{person}{Joshua Hibschman}, \bibinfo{person}{Daniel Afergan}, \bibinfo{person}{Yang Li}, \bibinfo{person}{Jeffrey Nichols}, {and} \bibinfo{person}{Ranjitha Kumar}.} \bibinfo{year}{2017}\natexlab{}.
\newblock \showarticletitle{Rico: A mobile app dataset for building data-driven design applications}. In \bibinfo{booktitle}{\emph{Proceedings of the 30th annual ACM symposium on user interface software and technology}}. \bibinfo{pages}{845--854}.
\newblock


\bibitem[Feiz et~al\mbox{.}(2022)]%
        {feiz2022understanding}
\bibfield{author}{\bibinfo{person}{Shirin Feiz}, \bibinfo{person}{Jason Wu}, \bibinfo{person}{Xiaoyi Zhang}, \bibinfo{person}{Amanda Swearngin}, \bibinfo{person}{Titus Barik}, {and} \bibinfo{person}{Jeffrey Nichols}.} \bibinfo{year}{2022}\natexlab{}.
\newblock \showarticletitle{Understanding screen relationships from screenshots of smartphone applications}. In \bibinfo{booktitle}{\emph{27th International Conference on Intelligent User Interfaces}}. \bibinfo{pages}{447--458}.
\newblock


\bibitem[Google(2023)]%
        {android-emulator}
\bibfield{author}{\bibinfo{person}{Google}.} \bibinfo{year}{2023}\natexlab{}.
\newblock \bibinfo{title}{Run apps on the Android Emulator | Android Developers}.
\newblock \bibinfo{howpublished}{\url{https://developer.android.com/studio/run/emulator}}.
\newblock


\bibitem[Google(2024a)]%
        {android-webview}
\bibfield{author}{\bibinfo{person}{Google}.} \bibinfo{year}{2024}\natexlab{a}.
\newblock \bibinfo{title}{Build web apps in WebView}.
\newblock \bibinfo{howpublished}{\url{https://developer.android.com/develop/ui/views/layout/webapps/webview}}.
\newblock


\bibitem[Google(2024b)]%
        {google-assistant}
\bibfield{author}{\bibinfo{person}{Google}.} \bibinfo{year}{2024}\natexlab{b}.
\newblock \bibinfo{title}{Google Assistant, your own personal Google}.
\newblock \bibinfo{howpublished}{\url{https://www.apple.com/siri/}}.
\newblock


\bibitem[Hong et~al\mbox{.}(2023)]%
        {hong2023cogagent}
\bibfield{author}{\bibinfo{person}{Wenyi Hong}, \bibinfo{person}{Weihan Wang}, \bibinfo{person}{Qingsong Lv}, \bibinfo{person}{Jiazheng Xu}, \bibinfo{person}{Wenmeng Yu}, \bibinfo{person}{Junhui Ji}, \bibinfo{person}{Yan Wang}, \bibinfo{person}{Zihan Wang}, \bibinfo{person}{Yuxiao Dong}, \bibinfo{person}{Ming Ding}, {et~al\mbox{.}}} \bibinfo{year}{2023}\natexlab{}.
\newblock \showarticletitle{CogAgent: A Visual Language Model for GUI Agents}.
\newblock \bibinfo{journal}{\emph{arXiv preprint arXiv:2312.08914}} (\bibinfo{year}{2023}).
\newblock


\bibitem[Lee et~al\mbox{.}(2023)]%
        {lee2023explore}
\bibfield{author}{\bibinfo{person}{Sunjae Lee}, \bibinfo{person}{Junyoung Choi}, \bibinfo{person}{Jungjae Lee}, \bibinfo{person}{Hojun Choi}, \bibinfo{person}{Steven~Y Ko}, \bibinfo{person}{Sangeun Oh}, {and} \bibinfo{person}{Insik Shin}.} \bibinfo{year}{2023}\natexlab{}.
\newblock \showarticletitle{Explore, Select, Derive, and Recall: Augmenting LLM with Human-like Memory for Mobile Task Automation}.
\newblock \bibinfo{journal}{\emph{arXiv preprint arXiv:2312.03003}} (\bibinfo{year}{2023}).
\newblock


\bibitem[Li et~al\mbox{.}(2023)]%
        {li2023blip}
\bibfield{author}{\bibinfo{person}{Junnan Li}, \bibinfo{person}{Dongxu Li}, \bibinfo{person}{Silvio Savarese}, {and} \bibinfo{person}{Steven Hoi}.} \bibinfo{year}{2023}\natexlab{}.
\newblock \showarticletitle{Blip-2: Bootstrapping language-image pre-training with frozen image encoders and large language models}.
\newblock \bibinfo{journal}{\emph{arXiv preprint arXiv:2301.12597}} (\bibinfo{year}{2023}).
\newblock


\bibitem[Li et~al\mbox{.}(2020)]%
        {li2020mapping}
\bibfield{author}{\bibinfo{person}{Yang Li}, \bibinfo{person}{Jiacong He}, \bibinfo{person}{Xin Zhou}, \bibinfo{person}{Yuan Zhang}, {and} \bibinfo{person}{Jason Baldridge}.} \bibinfo{year}{2020}\natexlab{}.
\newblock \showarticletitle{Mapping natural language instructions to mobile UI action sequences}.
\newblock \bibinfo{journal}{\emph{arXiv preprint arXiv:2005.03776}} (\bibinfo{year}{2020}).
\newblock


\bibitem[Li et~al\mbox{.}(2024)]%
        {li2024personal}
\bibfield{author}{\bibinfo{person}{Yuanchun Li}, \bibinfo{person}{Hao Wen}, \bibinfo{person}{Weijun Wang}, \bibinfo{person}{Xiangyu Li}, \bibinfo{person}{Yizhen Yuan}, \bibinfo{person}{Guohong Liu}, \bibinfo{person}{Jiacheng Liu}, \bibinfo{person}{Wenxing Xu}, \bibinfo{person}{Xiang Wang}, \bibinfo{person}{Yi Sun}, {et~al\mbox{.}}} \bibinfo{year}{2024}\natexlab{}.
\newblock \showarticletitle{Personal LLM Agents: Insights and Survey about the Capability, Efficiency and Security}.
\newblock \bibinfo{journal}{\emph{arXiv preprint arXiv:2401.05459}} (\bibinfo{year}{2024}).
\newblock


\bibitem[Liu et~al\mbox{.}(2023)]%
        {liu2023visual}
\bibfield{author}{\bibinfo{person}{Haotian Liu}, \bibinfo{person}{Chunyuan Li}, \bibinfo{person}{Qingyang Wu}, {and} \bibinfo{person}{Yong~Jae Lee}.} \bibinfo{year}{2023}\natexlab{}.
\newblock \bibinfo{title}{Visual Instruction Tuning}.
\newblock
\newblock
\showeprint[arxiv]{2304.08485}~[cs.CV]


\bibitem[Ma et~al\mbox{.}(2024)]%
        {ma2024comprehensive}
\bibfield{author}{\bibinfo{person}{Xinbei Ma}, \bibinfo{person}{Zhuosheng Zhang}, {and} \bibinfo{person}{Hai Zhao}.} \bibinfo{year}{2024}\natexlab{}.
\newblock \showarticletitle{Comprehensive Cognitive LLM Agent for Smartphone GUI Automation}.
\newblock \bibinfo{journal}{\emph{arXiv preprint arXiv:2402.11941}} (\bibinfo{year}{2024}).
\newblock


\bibitem[OpenAI(2023)]%
        {gpt4v}
\bibfield{author}{\bibinfo{person}{OpenAI}.} \bibinfo{year}{2023}\natexlab{}.
\newblock \bibinfo{title}{GPT-4V(ision) system card}.
\newblock \bibinfo{howpublished}{\url{https://openai.com/research/gpt-4v-system-card}}.
\newblock


\bibitem[Pan et~al\mbox{.}(2023)]%
        {pan2023autotask}
\bibfield{author}{\bibinfo{person}{Lihang Pan}, \bibinfo{person}{Bowen Wang}, \bibinfo{person}{Chun Yu}, \bibinfo{person}{Yuxuan Chen}, \bibinfo{person}{Xiangyu Zhang}, {and} \bibinfo{person}{Yuanchun Shi}.} \bibinfo{year}{2023}\natexlab{}.
\newblock \showarticletitle{AutoTask: Executing Arbitrary Voice Commands by Exploring and Learning from Mobile GUI}.
\newblock \bibinfo{journal}{\emph{arXiv preprint arXiv:2312.16062}} (\bibinfo{year}{2023}).
\newblock


\bibitem[Raffel et~al\mbox{.}(2020)]%
        {raffel2020exploring}
\bibfield{author}{\bibinfo{person}{Colin Raffel}, \bibinfo{person}{Noam Shazeer}, \bibinfo{person}{Adam Roberts}, \bibinfo{person}{Katherine Lee}, \bibinfo{person}{Sharan Narang}, \bibinfo{person}{Michael Matena}, \bibinfo{person}{Yanqi Zhou}, \bibinfo{person}{Wei Li}, {and} \bibinfo{person}{Peter~J Liu}.} \bibinfo{year}{2020}\natexlab{}.
\newblock \showarticletitle{Exploring the limits of transfer learning with a unified text-to-text transformer}.
\newblock \bibinfo{journal}{\emph{The Journal of Machine Learning Research}} \bibinfo{volume}{21}, \bibinfo{number}{1} (\bibinfo{year}{2020}), \bibinfo{pages}{5485--5551}.
\newblock


\bibitem[Rawles et~al\mbox{.}(2023)]%
        {rawles2023android}
\bibfield{author}{\bibinfo{person}{Christopher Rawles}, \bibinfo{person}{Alice Li}, \bibinfo{person}{Daniel Rodriguez}, \bibinfo{person}{Oriana Riva}, {and} \bibinfo{person}{Timothy Lillicrap}.} \bibinfo{year}{2023}\natexlab{}.
\newblock \showarticletitle{Android in the wild: A large-scale dataset for android device control}.
\newblock \bibinfo{journal}{\emph{arXiv preprint arXiv:2307.10088}} (\bibinfo{year}{2023}).
\newblock


\bibitem[Sun et~al\mbox{.}(2022)]%
        {sun2022meta}
\bibfield{author}{\bibinfo{person}{Liangtai Sun}, \bibinfo{person}{Xingyu Chen}, \bibinfo{person}{Lu Chen}, \bibinfo{person}{Tianle Dai}, \bibinfo{person}{Zichen Zhu}, {and} \bibinfo{person}{Kai Yu}.} \bibinfo{year}{2022}\natexlab{}.
\newblock \showarticletitle{META-GUI: Towards Multi-modal Conversational Agents on Mobile GUI}.
\newblock \bibinfo{journal}{\emph{arXiv preprint arXiv:2205.11029}} (\bibinfo{year}{2022}).
\newblock


\bibitem[Taeb et~al\mbox{.}(2023)]%
        {taeb2023axnav}
\bibfield{author}{\bibinfo{person}{Maryam Taeb}, \bibinfo{person}{Amanda Swearngin}, \bibinfo{person}{Eldon School}, \bibinfo{person}{Ruijia Cheng}, \bibinfo{person}{Yue Jiang}, {and} \bibinfo{person}{Jeffrey Nichols}.} \bibinfo{year}{2023}\natexlab{}.
\newblock \showarticletitle{AXNav: Replaying Accessibility Tests from Natural Language}.
\newblock \bibinfo{journal}{\emph{arXiv preprint arXiv:2310.02424}} (\bibinfo{year}{2023}).
\newblock


\bibitem[Toyama et~al\mbox{.}(2021)]%
        {toyama2021androidenv}
\bibfield{author}{\bibinfo{person}{Daniel Toyama}, \bibinfo{person}{Philippe Hamel}, \bibinfo{person}{Anita Gergely}, \bibinfo{person}{Gheorghe Comanici}, \bibinfo{person}{Amelia Glaese}, \bibinfo{person}{Zafarali Ahmed}, \bibinfo{person}{Tyler Jackson}, \bibinfo{person}{Shibl Mourad}, {and} \bibinfo{person}{Doina Precup}.} \bibinfo{year}{2021}\natexlab{}.
\newblock \showarticletitle{Androidenv: A reinforcement learning platform for android}.
\newblock \bibinfo{journal}{\emph{arXiv preprint arXiv:2105.13231}} (\bibinfo{year}{2021}).
\newblock


\bibitem[Wang et~al\mbox{.}(2023)]%
        {wang2023enabling}
\bibfield{author}{\bibinfo{person}{Bryan Wang}, \bibinfo{person}{Gang Li}, {and} \bibinfo{person}{Yang Li}.} \bibinfo{year}{2023}\natexlab{}.
\newblock \showarticletitle{Enabling conversational interaction with mobile ui using large language models}. In \bibinfo{booktitle}{\emph{Proceedings of the 2023 CHI Conference on Human Factors in Computing Systems}}. \bibinfo{pages}{1--17}.
\newblock


\bibitem[Wen et~al\mbox{.}(2024)]%
        {wen2023empowering}
\bibfield{author}{\bibinfo{person}{Hao Wen}, \bibinfo{person}{Yuanchun Li}, \bibinfo{person}{Guohong Liu}, \bibinfo{person}{Shanhui Zhao}, \bibinfo{person}{Tao Yu}, \bibinfo{person}{Toby Jia-Jun Li}, \bibinfo{person}{Shiqi Jiang}, \bibinfo{person}{Yunhao Liu}, \bibinfo{person}{Yaqin Zhang}, {and} \bibinfo{person}{Yunxin Liu}.} \bibinfo{year}{2024}\natexlab{}.
\newblock \showarticletitle{Autodroid: Llm-powered task automation in android}.
\newblock  (\bibinfo{year}{2024}), \bibinfo{pages}{543--557}.
\newblock


\bibitem[Wu et~al\mbox{.}(2023a)]%
        {wu2023never}
\bibfield{author}{\bibinfo{person}{Jason Wu}, \bibinfo{person}{Rebecca Krosnick}, \bibinfo{person}{Eldon Schoop}, \bibinfo{person}{Amanda Swearngin}, \bibinfo{person}{Jeffrey~P Bigham}, {and} \bibinfo{person}{Jeffrey Nichols}.} \bibinfo{year}{2023}\natexlab{a}.
\newblock \showarticletitle{Never-ending Learning of User Interfaces}. In \bibinfo{booktitle}{\emph{Proceedings of the 36th Annual ACM Symposium on User Interface Software and Technology}}. \bibinfo{pages}{1--13}.
\newblock


\bibitem[Wu et~al\mbox{.}(2023b)]%
        {wu2023webui}
\bibfield{author}{\bibinfo{person}{Jason Wu}, \bibinfo{person}{Siyan Wang}, \bibinfo{person}{Siman Shen}, \bibinfo{person}{Yi-Hao Peng}, \bibinfo{person}{Jeffrey Nichols}, {and} \bibinfo{person}{Jeffrey~P Bigham}.} \bibinfo{year}{2023}\natexlab{b}.
\newblock \showarticletitle{WebUI: A Dataset for Enhancing Visual UI Understanding with Web Semantics}. In \bibinfo{booktitle}{\emph{Proceedings of the 2023 CHI Conference on Human Factors in Computing Systems}}. \bibinfo{pages}{1--14}.
\newblock


\bibitem[Xie et~al\mbox{.}(2020)]%
        {xie2020uied}
\bibfield{author}{\bibinfo{person}{Mulong Xie}, \bibinfo{person}{Sidong Feng}, \bibinfo{person}{Zhenchang Xing}, \bibinfo{person}{Jieshan Chen}, {and} \bibinfo{person}{Chunyang Chen}.} \bibinfo{year}{2020}\natexlab{}.
\newblock \showarticletitle{UIED: a hybrid tool for GUI element detection}. In \bibinfo{booktitle}{\emph{Proceedings of the 28th ACM Joint Meeting on European Software Engineering Conference and Symposium on the Foundations of Software Engineering}}. \bibinfo{pages}{1655--1659}.
\newblock


\bibitem[Xing et~al\mbox{.}(2024)]%
        {xing2024understanding}
\bibfield{author}{\bibinfo{person}{Mingzhe Xing}, \bibinfo{person}{Rongkai Zhang}, \bibinfo{person}{Hui Xue}, \bibinfo{person}{Qi Chen}, \bibinfo{person}{Fan Yang}, {and} \bibinfo{person}{Zhen Xiao}.} \bibinfo{year}{2024}\natexlab{}.
\newblock \showarticletitle{Understanding the Weakness of Large Language Model Agents within a Complex Android Environment}.
\newblock \bibinfo{journal}{\emph{arXiv preprint arXiv:2402.06596}} (\bibinfo{year}{2024}).
\newblock


\bibitem[Yan et~al\mbox{.}(2023)]%
        {yan2023gpt}
\bibfield{author}{\bibinfo{person}{An Yan}, \bibinfo{person}{Zhengyuan Yang}, \bibinfo{person}{Wanrong Zhu}, \bibinfo{person}{Kevin Lin}, \bibinfo{person}{Linjie Li}, \bibinfo{person}{Jianfeng Wang}, \bibinfo{person}{Jianwei Yang}, \bibinfo{person}{Yiwu Zhong}, \bibinfo{person}{Julian McAuley}, \bibinfo{person}{Jianfeng Gao}, {et~al\mbox{.}}} \bibinfo{year}{2023}\natexlab{}.
\newblock \showarticletitle{GPT-4V in Wonderland: Large Multimodal Models for Zero-Shot Smartphone GUI Navigation}.
\newblock \bibinfo{journal}{\emph{arXiv preprint arXiv:2311.07562}} (\bibinfo{year}{2023}).
\newblock


\bibitem[Yang et~al\mbox{.}(2023)]%
        {yang2023appagent}
\bibfield{author}{\bibinfo{person}{Zhao Yang}, \bibinfo{person}{Jiaxuan Liu}, \bibinfo{person}{Yucheng Han}, \bibinfo{person}{Xin Chen}, \bibinfo{person}{Zebiao Huang}, \bibinfo{person}{Bin Fu}, {and} \bibinfo{person}{Gang Yu}.} \bibinfo{year}{2023}\natexlab{}.
\newblock \showarticletitle{AppAgent: Multimodal Agents as Smartphone Users}.
\newblock \bibinfo{journal}{\emph{arXiv preprint arXiv:2312.13771}} (\bibinfo{year}{2023}).
\newblock


\bibitem[Zhan and Zhang(2023)]%
        {zhan2023you}
\bibfield{author}{\bibinfo{person}{Zhuosheng Zhan} {and} \bibinfo{person}{Aston Zhang}.} \bibinfo{year}{2023}\natexlab{}.
\newblock \showarticletitle{You only look at screens: Multimodal chain-of-action agents}.
\newblock \bibinfo{journal}{\emph{arXiv preprint arXiv:2309.11436}} (\bibinfo{year}{2023}).
\newblock


\bibitem[Zhang et~al\mbox{.}(2023)]%
        {zhang2023mobile}
\bibfield{author}{\bibinfo{person}{Danyang Zhang}, \bibinfo{person}{Lu Chen}, {and} \bibinfo{person}{Kai Yu}.} \bibinfo{year}{2023}\natexlab{}.
\newblock \showarticletitle{Mobile-env: A universal platform for training and evaluation of mobile interaction}.
\newblock \bibinfo{journal}{\emph{arXiv preprint arXiv:2305.08144}} (\bibinfo{year}{2023}).
\newblock


\bibitem[Zhou et~al\mbox{.}(2023)]%
        {zhou2023webarena}
\bibfield{author}{\bibinfo{person}{Shuyan Zhou}, \bibinfo{person}{Frank~F Xu}, \bibinfo{person}{Hao Zhu}, \bibinfo{person}{Xuhui Zhou}, \bibinfo{person}{Robert Lo}, \bibinfo{person}{Abishek Sridhar}, \bibinfo{person}{Xianyi Cheng}, \bibinfo{person}{Yonatan Bisk}, \bibinfo{person}{Daniel Fried}, \bibinfo{person}{Uri Alon}, {et~al\mbox{.}}} \bibinfo{year}{2023}\natexlab{}.
\newblock \showarticletitle{Webarena: A realistic web environment for building autonomous agents}.
\newblock \bibinfo{journal}{\emph{arXiv preprint arXiv:2307.13854}} (\bibinfo{year}{2023}).
\newblock


\end{thebibliography}
\bibliographystyle{ACM-Reference-Format}
\appendix\onecolumn
\newpage
\section{Appendix}

\subsection{APIs Provided by \agentenv}
\label{appendix:apis}

\sys provides three categories of APIs for mobile agent integration and UI automation task execution.
APIs, parameters, and their return values are listed in Table~\ref{tab:apis}.

\begin{table*}[h]
    \caption[]{APIs provided by \sys to mobile agents for UI automation task execution on real mobile devices.}
    \scalebox{0.9}{
    \begin{tabular}{|l|l|l|l|}
    \hline
    \textbf{API Category}           & \textbf{API}           & \textbf{Parameter}                                                                                                                                                                                                                           & \textbf{Return Value}                                                                                        \\ \hline
    Metadata query                  & get\_task\_instruction & None                                                                                                                                                                                                                                         & str: A task description in natural language.                                                                 \\ \hline
    \multirow{2}{*}{UI state query} & get\_screenshot        & None                                                                                                                                                                                                                                         & \begin{tabular}[c]{@{}l@{}}str: A base64 encoded string \\ representing the current screenshot.\end{tabular} \\ \cline{2-4} 
                                    & get\_view\_hierarchy   & None                                                                                                                                                                                                                                         & \begin{tabular}[c]{@{}l@{}}str: A string representing the\\ textual view hierarchy in XML.\end{tabular}      \\ \hline
    \multirow{7}{*}{Action space}   & post\_task\_complete   & None                                                                                                                                                                                                                                         & None                                                                                                         \\ \cline{2-4} 
                                    & post\_task\_impossible & None                                                                                                                                                                                                                                         & None                                                                                                         \\ \cline{2-4} 
                                    & post\_press\_home      & None                                                                                                                                                                                                                                         & None                                                                                                         \\ \cline{2-4} 
                                    & post\_press\_back      & None                                                                                                                                                                                                                                         & None                                                                                                         \\ \cline{2-4} 
                                    & post\_click            & \begin{tabular}[c]{@{}l@{}}x, y: Normalized x/y coordinates for a \\ targeted screen position.\end{tabular}                                                                                                                                  & None                                                                                                         \\ \cline{2-4} 
                                    & post\_type             & str: The text to be input.                                                                                                                                                                                                                   & None                                                                                                         \\ \cline{2-4} 
                                    & post\_swipe            & \begin{tabular}[c]{@{}l@{}}touch\_x, touch\_y: Normalized x/y \\ coordinates where the swipe begins.\\ lift\_x, lift\_y: Normalized x and y \\ coordinates where the swipe ends.\\ duration: The interval of the swipe gesture.\end{tabular} & None                                                                                                         \\ \hline
    \end{tabular}
    }
    \label{tab:apis}
    \end{table*}

\end{document}